%% file: ms.tex
\documentclass [11pt,preprint]{aastex}

\usepackage{apjfonts}
\usepackage{lscape}
\usepackage{amsmath}

\input{defs.tex}

\shortauthors{Green}  

\shorttitle{Carbon Stars in the SDSS}

\begin{document}

\slugcomment{ApJ, accepted, January 15, 2013} 

\title{Innocent Bystanders: Carbon Stars from the Sloan Digital Sky Survey} 

\author{Paul Green\altaffilmark{1}}
\altaffiltext{1}{Harvard-Smithsonian Center for Astrophysics,
  Cambridge, MA 02138}

\begin{abstract}
Among stars showing carbon molecular bands (C stars), the main
sequence dwarfs, likely in post-mass transfer binaries, are numerically
dominant in the Galaxy.  Via spectroscopic selection from the Sloan
Digital Sky Survey, we retrieve 1220 C stars, $\sim$5 times more than
previously known, including a wider variety than past techniques such
as color or grism selection have netted, and additionally yielding 167
DQ white dwarfs.  Of the C stars with proper motion measurements, we
identify 69\% as clearly dwarfs (dCs), while $\sim$7\% are
giants. The dCs likely span absolute magnitudes $M_i$ from $\sim$6.5
to 10.5.  ``G-type'' dC stars with weak CN and relatively blue colors
are probably the most massive dCs still cool enough to show C$_2$
bands.  We report Balmer emission in 22 dCs, none of which are
G-types.  We find 8 new DA/dC stars in composite spectrum binaries,
quadrupling the total sample of these ``smoking guns'' for AGB binary
mass transfer.  Eleven very red C stars with strong red CN bands
appear to be ``N''-type AGB stars at large Galactocentric distances,
one likely a new discovery in the dIrr galaxy Leo~A.  Two such stars
within 30\arcmin\, of each other may trace a previously unidentified
dwarf galaxy or tidal stream at $\sim$40\,kpc.  We explore the
multiwavelength properties of the sample and report the first X-ray
detection of a dC star, which shows strong Balmer emission.  Our own
spectroscopic survey additionally provides the dC surface density from
a complete sample of dwarfs limited by magnitude, color, and
proper-motion. 
\end{abstract}


\clearpage
\section{The Carbon Star Menagerie}      \label{sec:intro}

Carbon (C) stars are defined loosely as those showing molecular
absorption bands of carbon - C$_2$, CN or CH in their optical spectra
\citep{Secchi69}. The key difference between C and M-type stars is
their relative photospheric abundance of carbon and oxygen.  In M
stars, carbon in the stellar atmosphere is largely consumed by CO, but
there's plenty of oxygen left for molecular species like TiO, which
dominate M star spectra.  C stars, by contrast, have such high C/O
ratios that CO binds all the oxygen, with plenty of carbon remaining to
form C$_2$, CN or CH molecular bands.

As far as we know, nearly all carbon in stellar atmospheres
is produced by the triple-$\alpha$ process of helium fusion
(3\,$^4$He\,$\rightarrow\,^{12}$C) in the interiors of red giant 
stars.  C-rich material can only be dredged to the surface by episodes
of strong convection associated with shell He-burning episodes (the
so-called third dredge-up) that take place in asymptotic giant branch
(AGB) stars.  However, quite a variety of carbon-rich giant stars has
emerged.  C star classifications are well-described in \citet{Gray09}.  
Carbon giants are quickly identifiable by their strong bandheads, and
easily detected to large Galactocentric distances, so offer excellent
dynamical probes of the Galactic dark  matter potential  (e.g.,
\citealt{Mould85,Bothun91}).    

Bright ($R\lax$13) C stars were first classified in the Henry Draper
Catalog as N (red) and R (blue) type.  The N-type C stars are classic
AGB stars - they are luminous ($\overline{M_K}\sim-7$) and massive
$2\lax\msun\lax3$), with thin disk kinematics
\citep{Wall98,LloydEvans10}.  Many 
are strongly variable, with evidence for circumstellar dust.  Based on
their properties (abundances, photometry, kinematics, variability and
luminosity), most stars previously classified as late R-type stars are
probably misclassified N-type AGB carbon stars \citep{Zamora09}.  Many
early R stars have been incorrectly classified as well, and in reality
are classical CH stars (see below) or normal K giants.  {\em Bona
  fide} early R stars have $\overline{M_K}\sim-2$, which puts them
near the ``red clump'' of helium core-burning red giants.  They show
near solar metallicity, enhanced nitrogen, low $^{12}$C/$^{13}$C
ratios, no $s$-element enhancements, and normal binarity
\citep{McClure97}, so the origin of their atmospheric carbon remains
mysterious.

Less luminous C-rich giants with lower masses ($\lax\,2\msun$) include
the CH and Ba stars.  Both types show strong carbon and $s$-process
abundance enhancements, and neither type is found at AGB luminosities.
CH stars have high space velocities and halo kinematics, and 
are metal poor with [Fe/H] values ranging from -0.5 to -1.7  (e.g.,
\citealt{Vanture92}, \citealt{Wall98}, and references therein).  
Barium\,II (Ba) stars are generally earlier F-, G-, and K-type
objects, with young to old disk kinematics \citep{Lu91}.
Both CH and Ba giants show evidence for a high binary frequency from 
radial velocity variations \citep{McClure90}, with companions of
about 0.6\,\msun, which are presumably white dwarfs.  There is also
sparse UV photometric evidence for white dwarf (WD) companions to 
Ba giants \citep{Bohm-Vitense00}.  Ba dwarfs have significant far-UV
excess compared to a control sample of normal F-type dwarfs
\citet{Gray11}.  

The mounting evidence for white dwarf companions to CH and Ba stars,
and the theoretical expectation that carbon and $s$-process elements
are dredged up and expelled only by ``intrinsic'' carbon producers
(AGB stars), has led to the hypothesis that the enhanced abundances in
CH and Ba stars are ``extrinsic'', i.e., attributable to previous
episodes of mass transfer to the current primary from an AGB star,
which has since evolved into a white dwarf.  \citet{McClure97} studied
a sample of 10 subgiant CH stars, and concluded that they were all
post mass transfer binaries, with 900$<P<$4000d.  Given their binary
orbital properties, both CH and Ba stars \citep{Izzard10} are thought
to have suffered wind-accretion from a former AGB companion.  Other
extrinsic systems include the carbon-enhanced, s-process-rich, very
metal-poor (CEMP-s) stars, with periods $P$ from 200 - 2800d and
amplitudes $K<10$\kms (but one with $P$=3.4d and $K=52$\kms
\citep{Lucatello05}.  Carbon-enhanced metal-poor (CEMP) stars have an
overabundance of carbon with respect to iron ([C/Fe]$>+$1.0), and
represent a sizeable fraction of very metal-poor stars
([Fe/H]$<-$2.0).  Radial velocity studies reveal that those that are
$s$-process-rich (CEMP-s stars) are the metal-poor analogs of
classical CH stars \citep{Lucatello05,Izzard09}.

The discovery of the first {\em main sequence} carbon (dC) star G77-61
by \citet{Dahn77} was at the time quite surprising, but it made clear that
extrinsic processes can produce enhanced atmospheric abundances.  The
prototype dC G77-61 is a single-line spectroscopic binary showing no
emission lines, and the companion, invisible in the optical range, is a
WD with \Teff$<$ 6000\,K \citep{Dearborn86}.  The hypothesis was that
in a mass transfer binary (MTB), carbon rich material may be lost in a
stellar wind from a thermally-pulsing AGB star and accreted by its
companion main-sequence star.  In the current epoch, the ``polluted''
companion, as an innocent bystander, could be either dwarf or giant,
but the AGB donor has evolved to a white dwarf.  This MTB scenario was
strongly supported by the discovery  additional dC stars recognized by
their large proper motions during a search for faint high latitude
carbon (FHLC) stars \citep{Green91}.

Dwarf carbon stars are uniquely valuable systems, because 
even without any of the above signatures of binarity, they are all easily
identifiable as post-MTB systems simply by their \ctwo\, and CN bands.
Furthermore, the properties of dCs may help us understand much about
the amount of abundance-enhanced mass transferred, and how mixing
occurs in the recipient dC.  Depending on the host system properties, a
current dC star may have inherited a significant portion of its mass
(a third or more) by accretion, changing its overall structure in the
process. 

We have very little idea how common dCs really are.  What are their
space densities locally?  Are they predominantly disk, thick disk, or
spheroid?  The relatively high temperature of the white dwarfs in known
DA/dCs suggest an origin in the Galactic disk.  Cooler DAs are
likely much more prevalent, but simply harder to detect in the
presence of a main sequence dC. In more metal-poor populations, less
C-enhanced mass transfer is required to create C/O$>$1 atmospheres. 
G77-61 is itself one of the lowest metallicity stars known
at [Fe/H]=-4 \citep{Plez05}, so it could either be part
of the elusive Population\,III stars formed very early in the
galaxy, or perhaps more likely, its abundances have been affected by
some process during its companion's binary evolution that selectively 
removed metals from the system (e.g., \citealt{Waelkens91}).



Red dwarfs and red subdwarfs are low-mass main sequence stars.
Because these stars are numerically dominant in the Galaxy, and have
lifetimes exceeding a Hubble time \citep{Laughlin97}, they provide
superb tracers of the chemical and dynamical evolution of the Galaxy.
Only those examples that are carbon rich are easily identified from
low/intermediate resolution optical spectra as post-mass-transfer
binaries, and their present-day properties can provide a wealth of
information about the history of their former AGB companions and the
binary systems that they inhabit.  This makes dCs an important class
of object for study.




A large sample of FHLC stars with high quality multiwavelength data
clearly provides a myriad of research opportunities.  
Carbon stars are now known to span absolute magnitudes from 
late-type dwarfs ($M_V\sim 10$) to supergiants ($M_V\sim -4$).  
Luminosity discriminants from photometric colors and/or
low/intermediate-resolution spectroscopy, as yet elusive, would be
extremely valuable \citep{Green94,Downes04}.  Carbon giants can be
used as dynamical tracers of the Galactic potential (e.g.,
\citealt{Mould85,Bothun91}) or signals of dwarf galaxies or their
remnant tidal streams.
Dwarf carbon stars, expected to be in binary systems, should be all
radial velocity variable, yet the binarity and orbital parameters for
dCs as a population are completely unknown.  A large sample of bright
dCs should provide useful candidates for radial velocity monitoring.
The frequency of different mass transfer
scenarios for dCs, e.g., wind or Roche lobe overflow, can be modeled
and probed.  A reasonable sample of composite dC/WD systems can
constrain the age since mass transfer for those systems.
The space density of dCs can constrain the properties of ancient binary
systems - their mass ratios, separations, and evolution \citep{deKool95}.
Parallaxes can be obtained for some dCs, yielding more robust
information on the distribution of their absolute magnitudes
\citep{Harris98}.  Some of the brighter systems may soon be candidates
for detailed modeling and abundance analysis \citep{Plez05}, yielding
information about the production of elements in extinct generations of
AGB stars, dating back to the early Universe.\\

\section{SDSS Sample Selection} 
\label{sec:sample}

In this paper, we use the Sloan Digital Sky Survey (SDSS;
\citealt{York00}) to greatly expand the sample of confirmed FHLC
stars. Our goal is to select a large sample of C stars from the SDSS
with minimal pre-selection by color, so that a wider variety of C
stars can be studied than in the past.  Using the SDSS DR7 database,
after some experimentation, we chose to search for stars with spectra
showing strong cross-correlation coefficients with the SDSS carbon
star templates numbered 18 and 19, used for spectral cross-correlation
by the SDSS project.\footnote{See
  http://www.sdss.org/dr5/algorithms/spectemplates/. Template 20
  yielded too many hot white dwarfs and/or blue stars.}  SDSS
cross-correlation follows the technique of \citet{Tonry79}, where the
three highest cross-correlation function (CCF) peaks are found, fitted
with parabolas, and output with their associated confidence limits. We
required a best-fit peak CCF height of at least 0.3, which yielded
about 400 stars, of which less than 0.5\%, were not C stars.  On the
other hand, a minimum peak height of 0.1 yielded 8846 objects, few of
which were C stars.  After visual inspection of all spectra, we found
that a minimum peak height of 0.15 yielded almost entirely C
stars. The most common exceptions were quasars near $z=3.66$ (where
the blue wing of Ly$\alpha$ emission appears at the rest wavelength of
the $\lambda$5636 Swan band of C$_2$) and emission line or Seyfert
galaxies near $z=0.047$ (where the H$\beta$/[OIII] emission complex
appears at the rest wavelength of the $\lambda$5165 Swan band of
C$_2$).

We recovered 1223 {\em bona fide} C stars in this procedure, which
included 30 DQ stars.   There are 1209 stars that are well-matched to
template number 18, which has strong Swan bands.  Of these, 
422 also match with template 19, a redder template with strong red CN 
bands at $\lambda\lambda$7876, 7899.  Only 14 stars match with
template 19 only, including 5 stars that show at best weak evidence
for C$_2$ but which have strong red CN bands.  In some cases,
these stars are very red, so that while CN is evident, C$_2$ which may 
be intrinsically strong is lost in noise.  

From SDSS DR8, we selected objects with spectroscopic class
STAR and any subclass including 'Carbon', e.g., Carbon, Carbon\_lines,
or CarbonWD\footnote{Although our primary interest 
is FHLC stars, we include CarbonWD because the classification may not
always be correct.}, as long as the velocity was within 3,000 km/sec
of zero. The latter velocity criterion eliminates some QSOs
incorrectly identified as carbon stars.   Our selection initially yielded
1604 objects. 975 were in common with the DR7 velocity template
selection, while 629 were newly-identified objects. However,
after visual inspection, first we removed about 240 objects
because of low S/N, and then another 40 were removed as having
smooth blue continuum.  (Some of these may well be DQ white dwarfs.) 
In the end, our list included 1264 spectra with visually-confirmed
carbon molecular band features.

We merged the DR7 and DR8 lists and retrieved uniform DR8 data where
available (13 DR7 objects had none), yielding 1510 spectra.  There
are multiple spectra for some objects, so that the number of
distinct stars is 1390.  

\begin{figure}
\includegraphics[scale=0.65,angle=-90]{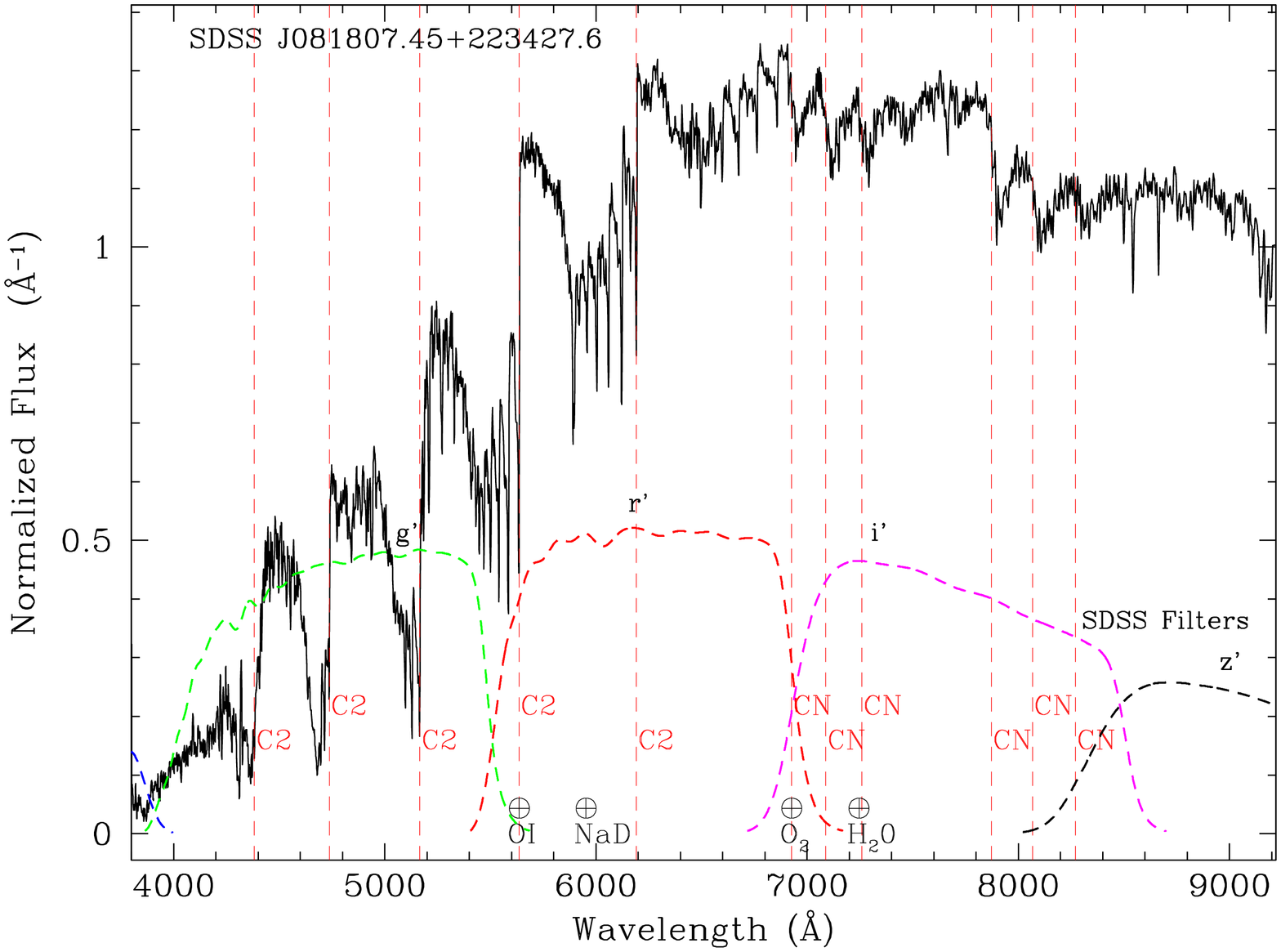}
\caption{Spectrum of a typical dC star. This SDSS spectrum of
SDSS\,J081807.45+223427.6, an extreme high proper motion dC star
(total proper motion $\mu$=246\,mas/year), shows typical \ctwo\, and
CN band heads marked with red dashed lines (wavelength list from
\citet{Davis87}). Approximate SDSS $griz$ filter transmission curves
are shown in dashed lines. Positions of strong potentially telluric
features are marked across the bottom.
\label{dCspec}}
\end{figure}

Generally, FHLC stars show a wide variety of colors, molecular
and line absorption features, of which examples can be seen
in \citet{Margon02}.  Figure\,\ref{dCspec} shows a typical
dC spectrum, for a high proper motion example newly reported here.
The spectrum is overlaid with SDSS filter curves, which serves to
illustrate the primary reason that C stars diverge from the
stellar locus - strong \ctwo\, molecular bands remove flux from the
$g$ bandpass. Our visual inspection of the spectra yielded just a
handful of principal classifications: these include ``normal'' dC
stars, 134 G-type FHLC stars,\footnote{The G-type FHLC stars are
not intrinsically distinct or well-separated, so this number
is a matter of definition - see \S\,\ref{sec:gtypes}.} 167 DQ White
Dwarf Stars, and 9 composite spectrum DA/dC systems.  Because they
contain degenerate stars and are not the focus of the current work, we
remove the latter two classes, leaving 1,211 objects in our main
sample in Table\,\ref{tab:fhlc}.   We discuss the principal
classifications further below.   

Comparing to the largest previous FHLC star compilation to date of
251 SDSS stars in \citet{Downes04}, our sample contains all but 
24 of their objects.  Most were rejected by our own visual inspection,
because they had (1) very low S/N that made a C star classification
doubtful, or (2) visual evidence in F- or G-type stars for strong
G-band of CH, but not for \ctwo\, or CN bands.  For completeness,
we list those objects in Table\,\ref{Downes04NotHere}.

\section{G-type FHLC Stars} 
\label{sec:gtypes}

\begin{figure}
\includegraphics[scale=0.65,angle=-90]{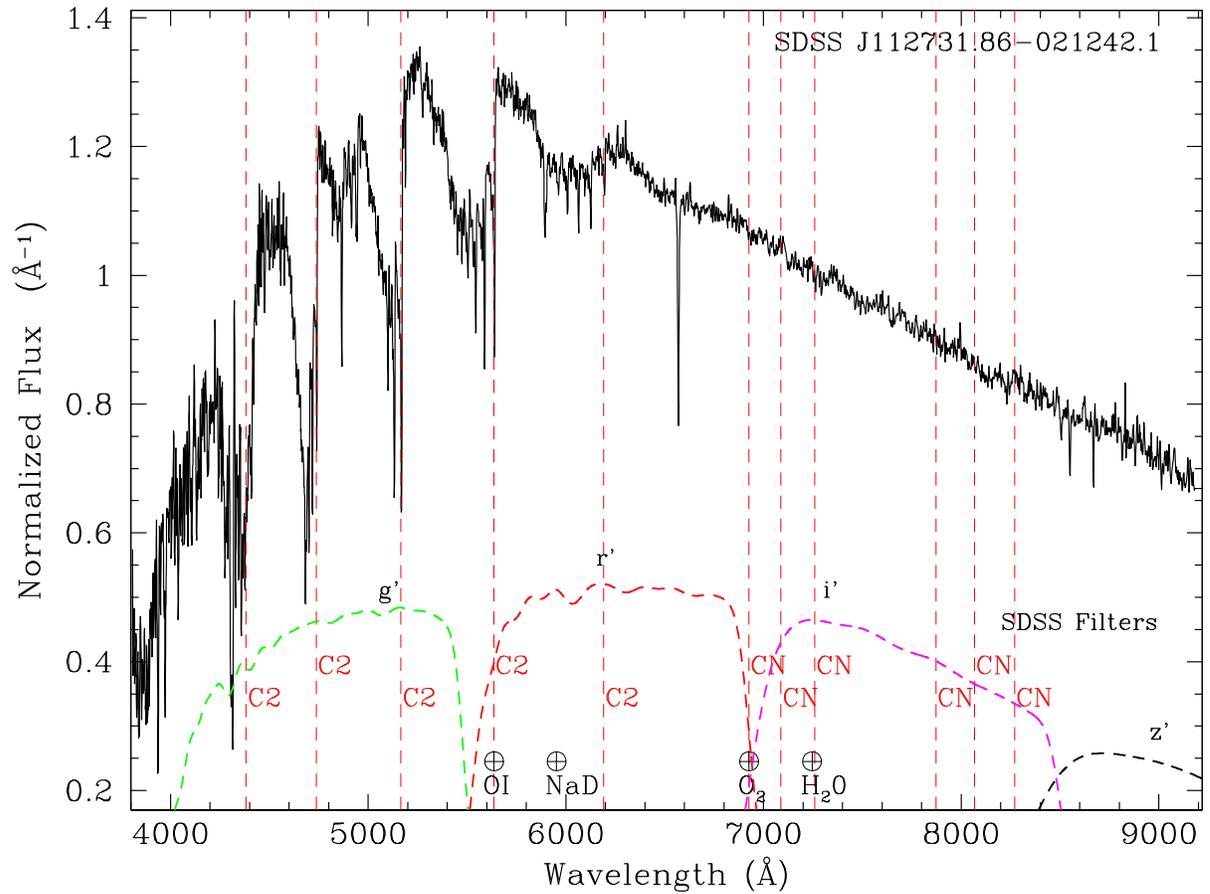}
\caption{Spectrum of a G-type FHLC star. This SDSS spectrum of
the G-type dC SDSS\,J112731.86-021242.1 ($\mu$=42\,mas/year) shows
features typical of the G-type C stars - blue continuum and 
no evidence for red CN bands.  
\label{Gtypespec}}
\end{figure}

One rather distinct group of C stars includes 134 objects with remarkably
similar spectra, typically having blue continuua, strong C$_2$ bands,
strong narrow Balmer and Ca absorption lines, but weak (usually
non-existent) red CN bands.  A typical example spectrum is shown in
Figure\,\ref{Gtypespec}. In the \umg\, vs. \gmr\, diagram of
Figure\,\ref{gmr_umg}, these objects, shown with open blue circles,
are strongly clumped.  In the \gmr\, vs. \rmi\, diagram\footnote{Their
  colors are not so well  distinguished in redder color-color
  diagrams.} of Figure\,\ref{rmi_gmr}, they hew close to the  
absolute mag locus of \citet{Kraus07}, alongside the thin-disk
sequence from about G2 - K0. For this reason, while \citet{Downes04}
referred to these as F/G carbon stars, we designate them simply as
G-type FHLC stars.  However, we note that the \ctwo\, bands generally
make FHLC stars appear bluer than similar-mass stars on the O-rich
main sequence.  The most likely absolute $i$-band magnitudes of G-type
dC stars ($M_i\sim7.25$; see \S\,\ref{sec:absmags}) puts them closer
to K7 - M0.

\section{DQ White Dwarf Stars} 
\label{sec:dqs}

Another distinct class that we select are the DQ stars, which are
white dwarfs showing atomic or molecular carbon features in their
spectra.  We find 167 such objects, listed in Table\,\ref{tab:dqs},
which have the expected high proper motions and blue colors.  The most
commonly-accepted explanation for the DQs is dredge-up from the
underlying carbon/oxygen core through the expanding He convection
zone, which explains stronger C features at higher effective
temperatures.  DQs have been reviewed by \citet{Dufour05}, and some 65
SDSS examples have been studied \citep{Koester06} spanning 5,000$<
T_{\rm eff}<$10,000\,K, with some hot DQs now known to extend up to
$\sim$24,000\,K \citep{Dufour08}.  Our DQ sample expands significantly
on the existing list of recognized DQs, especially for objects
with \gmr$>$0.5.  DQs can appear redder in \gmr\, because they are
cooler and/or lower [C/He] objects, but in our sample it is
due in every case to extremely strong Swan bands of \ctwo\, falling in
the $g$ bandpass.  Several of these DQs show molecular absorption
bands $\sim$100 -- 300\AA\, shortward of the \ctwo\, Swan bandheads.
These ``peculiar DQ bands'' were discussed recently by \citet{Hall08}
and \citet{Kowalski10}.  We do not discuss DQs extensively in this
paper, because they are degenerates, so neither C giants nor dwarf
stars, and are in any case thoroughly discussed elsewhere in the
literature.  However, 
the large number of DQs we find testifies to the sensitivity of our
methods for detecting even weak or pressure-broadened carbon features,
and for extending to objects that are significantly bluer than known C
stars. A handful of DQs at low S/N have strange continuum shapes
and/or evidence for CN molecular bands in the red, so may actually be
composite DA/dC stars.  

\section{Emission Line Stars} 
\label{sec:emlines}

Of our full sample of 1220 FHLC stars, 51 show clear Balmer emission
lines upon visual inspection.  Seven of those are DA/dC systems
(\S\,\ref{sec:dadcs}).  Of the remaining 44, 22 are definite dwarfs
(see \S\,\ref{sec:absmags}), 20 are of unknown luminosity class, and 2
are giants (including Draco C-1, a known symbiotic star in the Draco
dwarf galaxy).   None of the 134 G-type stars show emission lines.  
The 22 emission line dCs represent 3\% of the 724 definite dCs.
The 2 emission line giants represent about the same fraction of
giants, so emission lines do not at first glance appear to be a
luminosity diagnostic, even in a probabilistic sense. Further studies
of these emission line dCs are important, to determine the cause of
the emission.  Possibilities include coronal activity in the C star as
a result of spin-up from accretion and/or tidal locking.  Emission
lines caused by irradiation by a hot companion only seems likely for
the DA/dC systems.  

\section{Composite Spectrum DA/dC Stars} 
\label{sec:dadcs}

From our SDSS sample of C stars, a small but especially interesting
subsample includes DA/dC composite spectrum systems.  These show 
the usual clear \ctwo\, bands, often along with  red CN bands,
but their spectra display a strong blue continuum and the
pressure-broadened hydrogen Balmer absorption lines typical of a DA
white dwarf. There are only two such objects known to date
\citep{Heber93, Liebert94} \footnote{We disagree with \citet{Downes04}
  classification of SDSS\,J012747.73-100439.2 as a composite system.
}, but objects like these are the `smoking guns' that strongly support
the post binary mass transfer scenario for the origin of dCs.  We find
9 clear examples in our survey.  Only one of these obvious composite
systems, SDSS\,J151905.96+500702.9 is a known DA/dC system discussed
by \citet{Liebert94} as SBS\,1517+5017 or CBS\,311. Among our 8
newly-recognized DA/dC systems, all but two show Balmer emission lines
in their SDSS spectra.  At low S/N, it can be difficult to distinguish
between a DQ and a composite DA/dC system.  There are 3 such spectra
in this category, but none show emission lines, and we believe they
are likely to be {\em bona fide} DQ stars.

Because of the unique properties of DA/dC systems, there is widespread
interest in them beyond the community of carbon star afficionados,
including those involved in research of WDMS systems, close binaries,
CVs, and planetary nebulae, etc.  For this reason, we discuss the
DA/dC composite systems in a companion paper \citep{Green13}.

\section{Optical Colors} \label{sec:optcolors}

 Spectroscopic fibers are assigned to SDSS photometric objects 
based on diverse criteria which evolved significantly over the course
of the  survey.  Understanding the reasons why each object may have an
SDSS spectrum is crucial to understanding the biases in any SDSS
spectroscopic sample such as this one.  

Of the 1220 FHLCs in our sample (including the 8 DA/dCs), the majority
(56\%) were targeted as QSOs.  A quarter (24\%) were targeted by the
SEGUE project (objType={\tt NONLEGACY}), and just 17\% were targeted
as likely C stars using the color wedge from \citet{Margon02} shown in
Figure\,\ref{rmi_gmr}.  The remaining 2.6\% were targeted for a
variety of other reasons, including white dwarf candidates, galaxies,
reddening standards, and ROSAT counterparts (2).

Of the 134 G-type FHLCs, 97 (74\%) were targeted by SEGUE
based on optical colors, the bulk as low metallicity candidates.
So the large fraction of G-type stars in our sample compared to
\citet{Downes04} is due to different targeting by SEGUE. 

\begin{figure}
\plotone{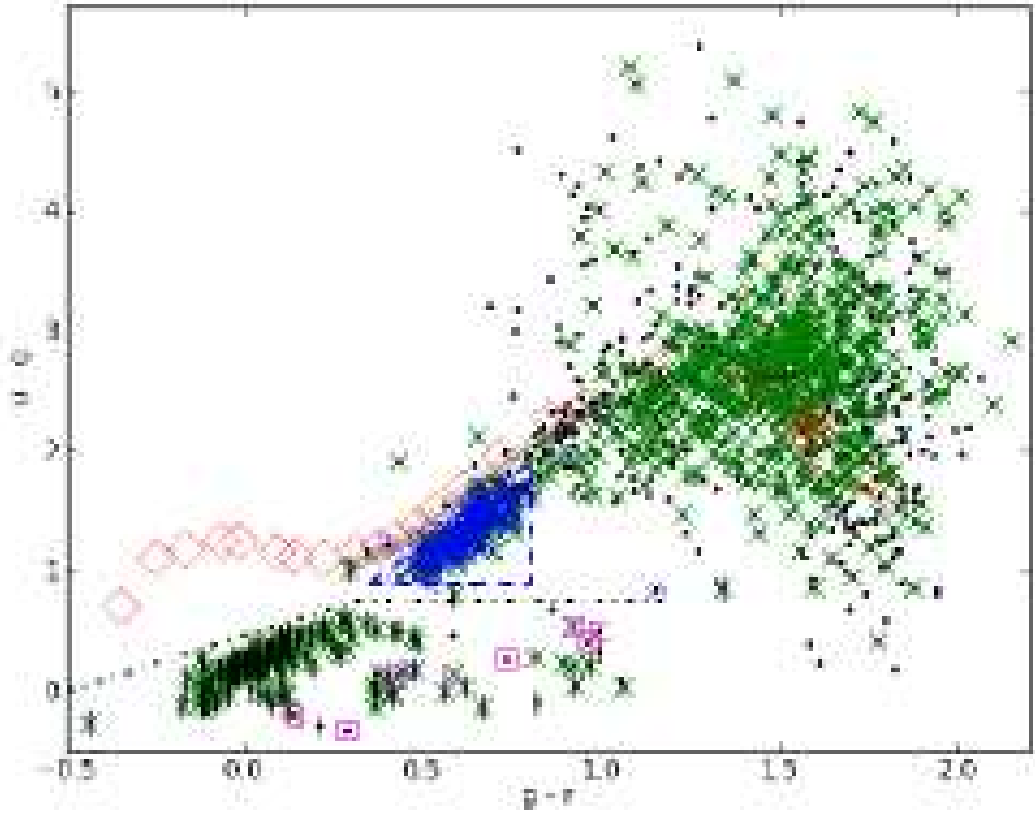}
\caption{Plot of \umg\, vs \gmr\, for SDSS stars in our sample
(black dots). Objects with significant proper motions (57\% of our
sample) have their symbols overlaid with a green cross.
DQ white dwarfs  (vertical bars) and composite DA+dC
systems (magenta boxes) are well segregated in this color space by the
black dot-dashed line. Blue circles mark the G-type FHLC stars, which
are efficiently delineated within the blue dashed triangle.  Large
red diamonds show colors from the stellar 
SED compilation of  \citet{Kraus07}.  The median and $\pm$25\% grey
contours are shown from the  SDSS stellar locus of \citet{Covey07},
for types with at least 200 stars in the sample.  
  \label{gmr_umg}}
\end{figure}

\begin{figure}
\plotone{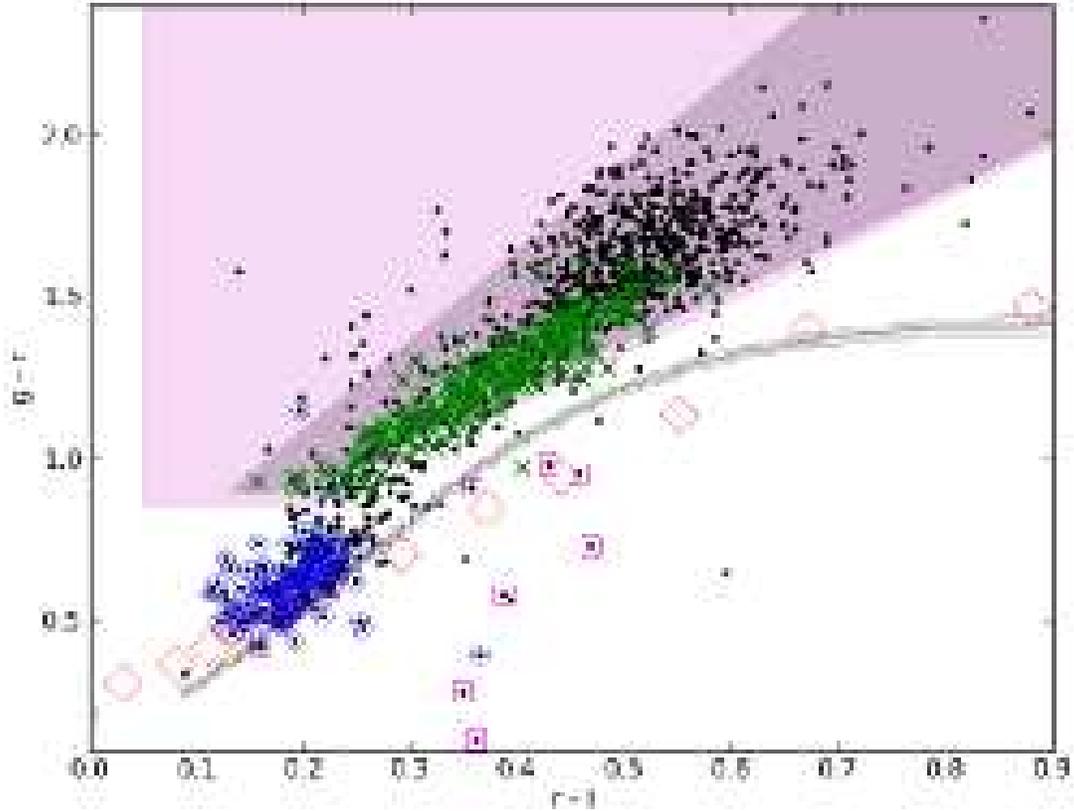}
\caption{Plot of \gmr\, vs \rmi\, for SDSS stars in our sample
(black dots).  Symbols are as described in Figure\,\ref{gmr_umg},
except that here the green crosses show objects targeted by SDSS
explicitly as carbon star candidates. Our spectroscopically-selected sample
significantly extends the color range of known dCs; only 17\% of our 
sample was targeted as FHLC candidates. DQs have also been removed as in
subsequent figures. The light magenta wedge shows   
the SDSS color selection suggested by \citet{Margon02}. The smaller
more deeply-shaded wedge shows the additional criterion used for our
FAST dC completeness study described in \S\ref{sec:FAST}. 
  \label{rmi_gmr}}
\end{figure}

Our FHLC stars span a wide range of colors in the \umg\, vs. \gmr\,
color diagram of Figure\,\ref{gmr_umg}.  For a handful of stars,
extreme colors are found resulting from saturation, or very faint $u$
band mags with large errors. However, several important clusterings
are in evidence.

DQ white dwarfs  and composite DA+dC systems are well segregated 
using $\umg<0.875\,\gmr + 0.5$ $\cup$  $\umg<$0.75. Stars
with blue \umg\, but \gmr$>$0.6 are either composite systems
or DQ\,pec stars with extremely deep Swan bands dimming
the $g$ bandpass.  

In Figure\,\ref{gmr_umg}, we plot large red diamonds to show colors
from the stellar SED compilation of
\citet{Kraus07}, which is most appropriate for O-rich main sequence
thin-disk, solar-metallicity 
stars.  Both C/O ratio and metallicity can strongly affect band
strengths and therefore colors, explaining the shift of the FHLCs away
from the main sequence disk locus of \citet{Covey07}, shown as a gray
line.  A clump of FHLC stars appears in a region offset just blueward of that, 
bordered by the following three lines shown by the blue dashed
triangle $\umg< 2\,\gmr + 0.3\,\cup$\,$\umg>0.9$\,$\cup$\,$\gmr<0.8$.  
But for a handful of exceptions, these $\sim$180 objects are all
the G-type FHLC stars described in \S\,\ref{sec:sample}.  Based on a much
smaller sample of these, \citep{Downes04} suggested that 
these G-type FHLC stars may form a continuum of properties with redder 
examples.  Here, the G-type FHLCs seem strongly clumped, 
and appear to show a gap separating them from the general FHLC
sequence (see also Figure\,\ref{rmi_gmr}).  However, this is largely the
effect of the SDSS fiber targeting algorithms.  The deeper
shaded region in Figure\,\ref{rmi_gmr} shows the color wedge
described in \citet{Margon02} which was used to target
potential FHLC stars for SDSS fibers.  Since our sample is selected
from SDSS spectra, FHLC stars within this color wedge are naturally 
over-represented relative to FHLCs elsewhere in the plane.

\begin{figure}
\plotone{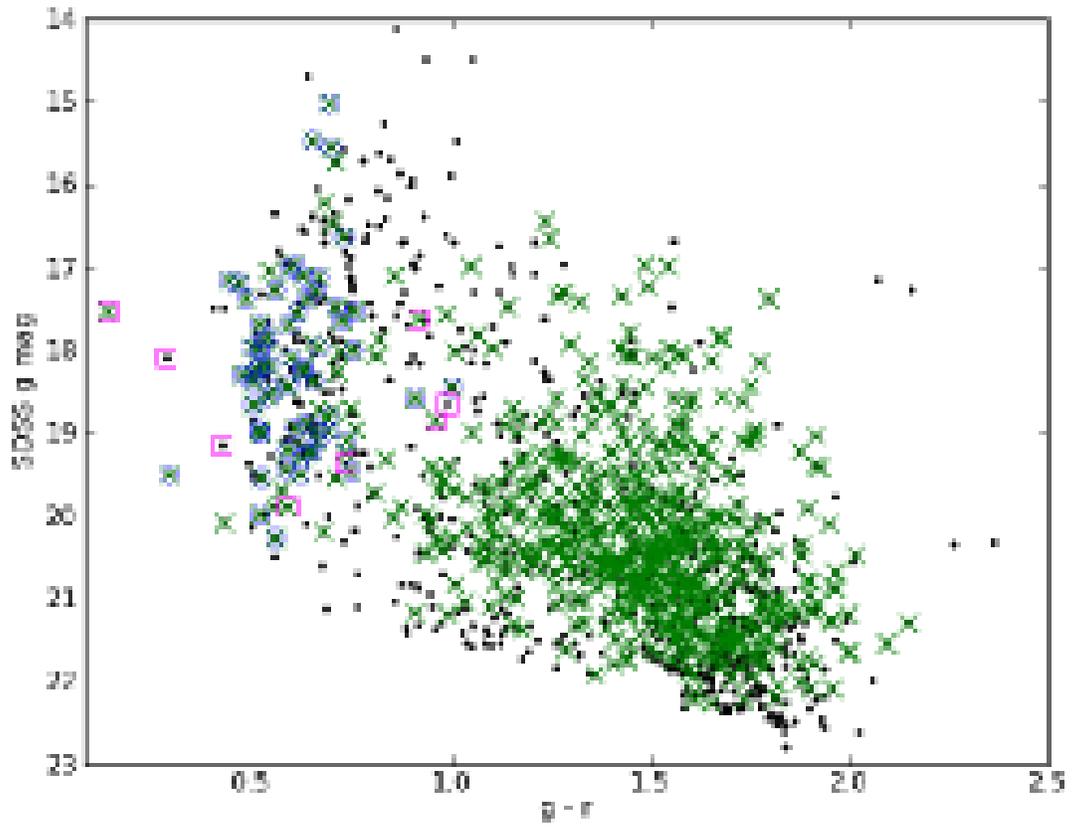}
\caption{Color-magnitude plot of $g$ magnitude vs \gmr\, for SDSS
C stars in our sample (black dots).  Symbols are as described in
Figure\,\ref{gmr_umg}. 
  \label{gmr_g}}
\end{figure}

In the $g$ vs \gmr\, color-magnitude diagram of Figure~\ref{gmr_g},
most FHLC stars, at \gmr$\sim$1.4 and faint magnitudes, coincide with
the locus of ``thick-disk-like'' ([Fe/H] $= -0.7$) stars seen at high
Galactic latitudes ($|b|>30\deg$) in the SDSS \citep{deJong10}.  The
brighter G-type FHLCs near \gmr$\sim 0.6$ instead coincide with the
main-sequence turn-off stars of older, thick-disk or
'inner-halo-like'' stars of intermediate metallicity 
([Fe/H]$ =-1.3$).  The carbon molecular bands make FHLC stars bluer in
\gmr\, making identification with these populations provisional at best.
However, CH giants, which by their binarity are known to also be
extrinsic, post-MTB systems, are also metal poor with [Fe/H] values
ranging from -0.5 to -1.7 (e.g., \citealt{Vanture92},
\citealt{Wall98}, and references therein).

We assume that all dCs are in post-mass transfer binaries.  Thus, they
are clearly in systems where the initial primary has already turned off
the main sequence, and indeed completed the AGB phase.  If the mass
ratio $q$ of binary systems peaks near unity \citep{Reggiani11}, then
most dCs should be close to the turnoff mass, especially if the mass
accretion is enhanced in higher mass bystanders.  However, above
a certain mass and effective temperature, the formation of \ctwo\,
molecules is suppressed, and at some point the \ctwo\, features by
which our FHLCs are identified diminish below detectability.
Higher mass objects might still be identified as carbon-enhanced metal
poor stars, but only with spectroscopy of significantly higher
resolution and S/N. 

\begin{figure}
\plottwo{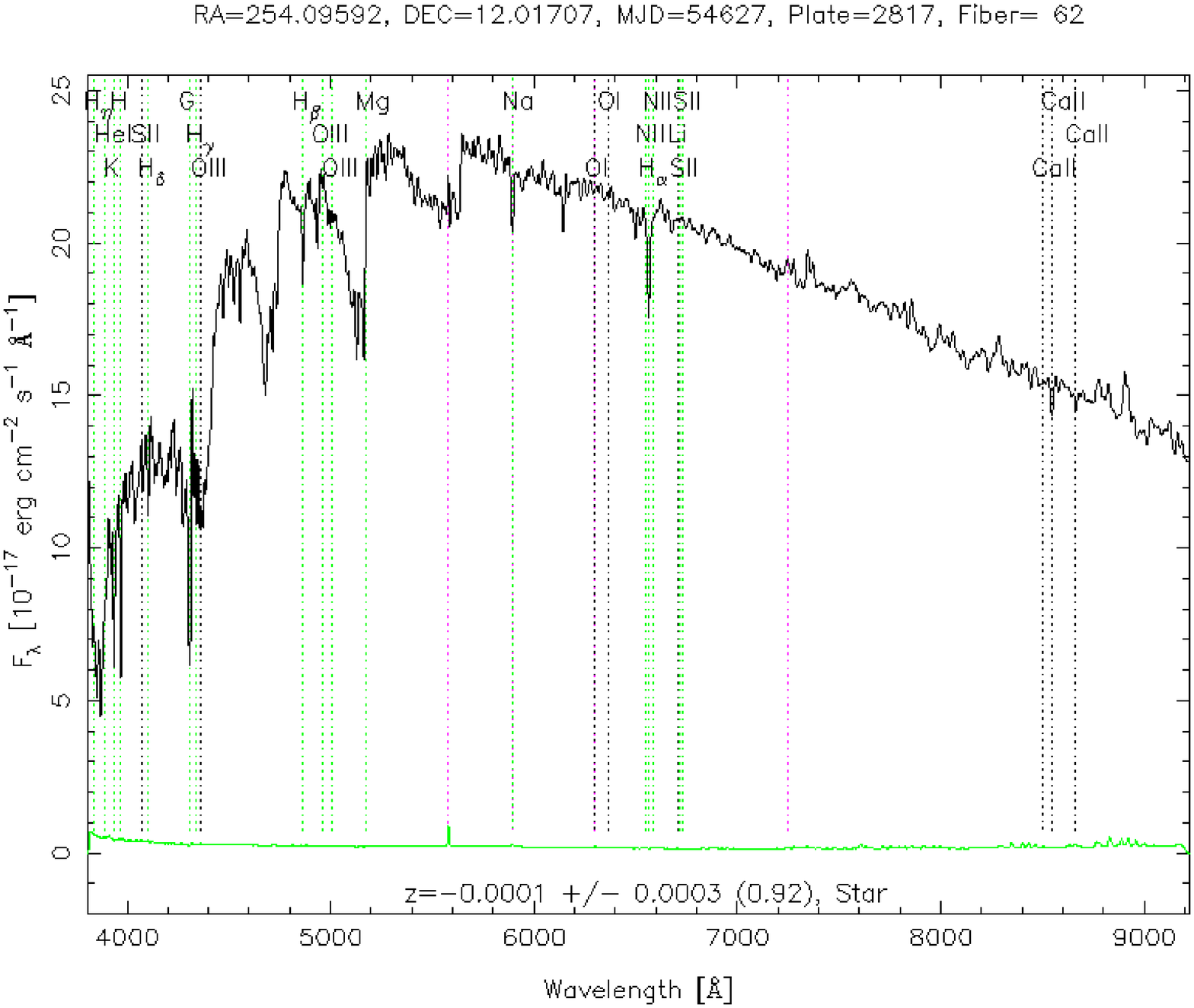}{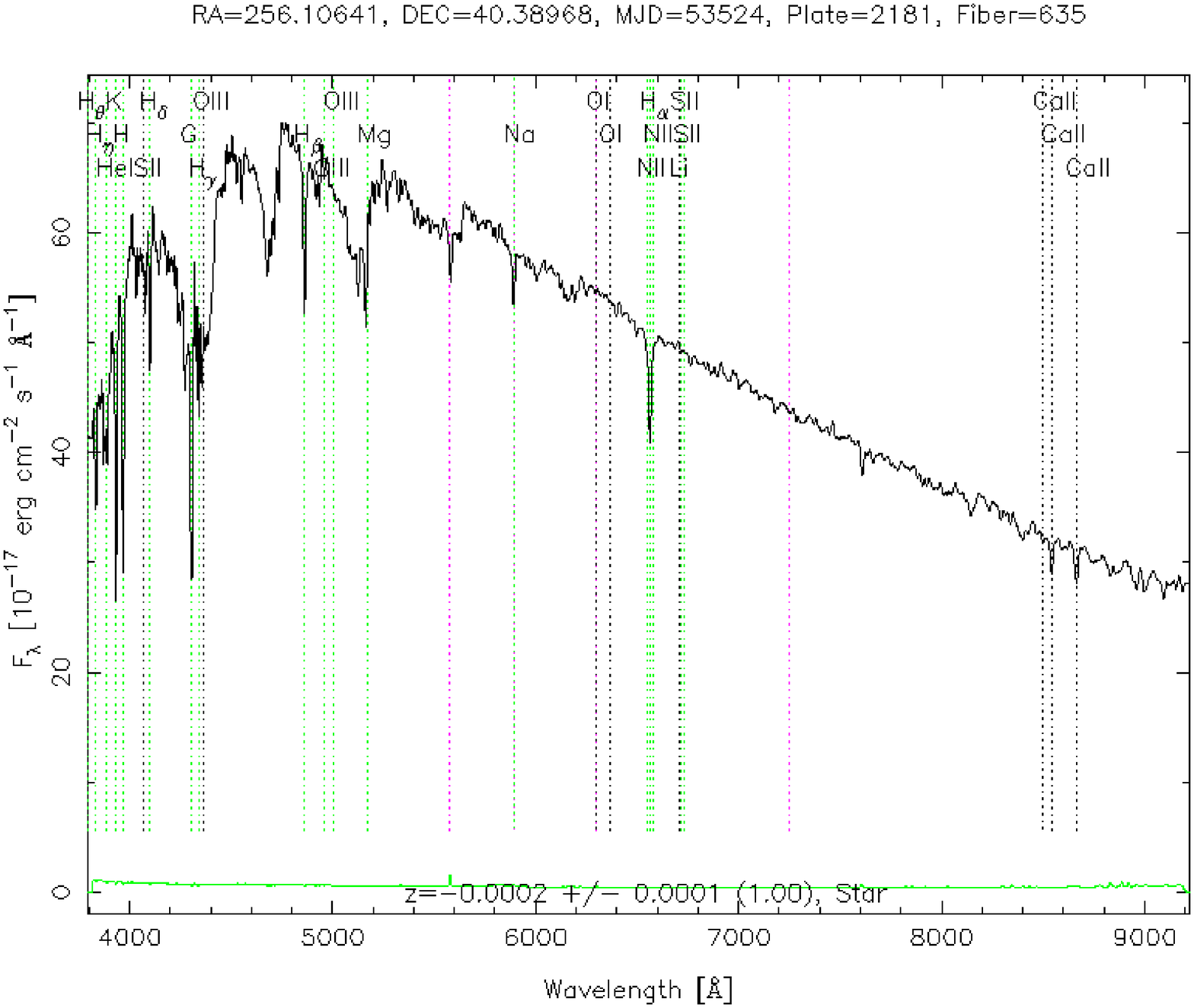}
\caption{Representative SDSS spectra of G-type stars.
G-type stars on the red edge of the clump in Figure\,\ref{rmi_gmr}
(e.g., with $\gmr>0.7$ and $0.2<\rmi<0.25$), 
like SDSS\,J165623.0+120101.4 
(left) have significantly stronger carbon molecular bands than do
those at the blue edge (e.g., $\gmr<0.5$ and $0.1<\rmi<0.15$), like
SDSS\,J170425.5+402322.8
(right). The lack of identified FHLC stars bluer than this edge may
simply be due to the disappearance of C$_2$ bandheads at warmer
temperatures. 
  \label{GtypeCbands}}
\end{figure}

We thus hypothesize that the current-day main sequence primaries
in post-AGB MTB systems should mostly be near the turnoff.  Because they
are more massive and brighter, among post-MTB survivors, such stars
should be preponderant in a magnitude-limited sample such as the SDSS.
At higher masses, they may also be more effective at accreting
abundance-enhanced material.  Since they are essentially innocent
bystanders to mass accretion, one might expect a smooth trend in color
and mass for these stars in Figure\,\ref{rmi_gmr}.  Instead, there
is a distinct clump of G-type FHLC stars, separated by a gap from
redder dCs.  This is probably caused by two effects.  First, the SDSS
color selection wedge for C star spectroscopic candidates causes the
cutoff of dCs near \gmr = 0.9.  If not for that color selection
wedge, we would expect a smooth trend of FHLC stars across the gap
that spans $0.8<\gmr <0.9$.  Second, as the G-type FHLCs become
progressively hotter and bluer, the carbon molecular bandheads weaken
(see Figure\,\ref{GtypeCbands}), until they are no longer detectable
for stars bluer than about \gmr\,=0.4. Several effects thus conspire
to make the G-type FHLC stars stand out in a clump in our sample.
Significant fractions of bluer stars may well have mass
transfer-enhanced abundances at masses higher than this, but more
detailed analysis of their spectra are required, probably at higher
resolution and S/N.

\begin{figure}
\plottwo{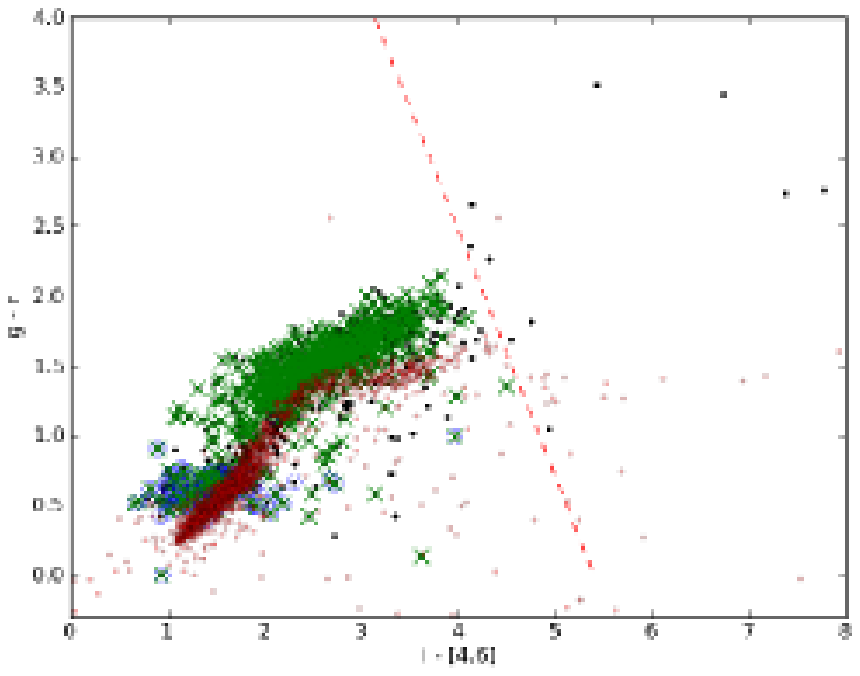}{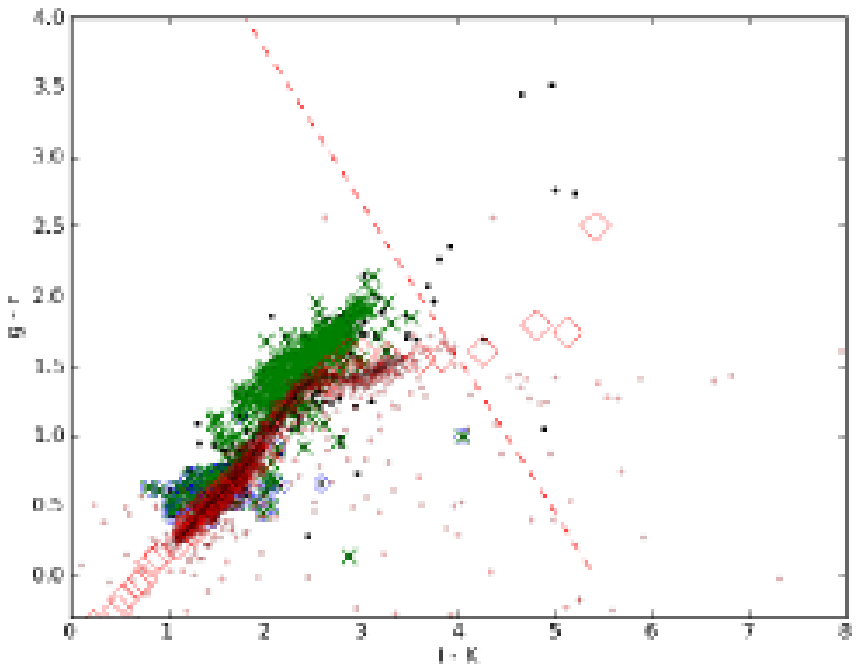}
\caption{LEFT: Plot of ($i$ - [4.6]) vs \gmr\, colors for SDSS FHLC stars in
our sample (black dots) detected by WISE.  G-type FHLC stars are shown
as open blue circles. Green crosses indicate objects with significant
proper motion.  Red circles are generic high proper motion dwarfs
($\mu>20$\,mas/yr) from the SDSS DR1/USNO-B catalog of \citet{Gould04}.  
The red dashed line at $(g-r)> -1.35\,(i-[4.6]) + 7.5$
shows an effective demarcation for AGB stars.  However, either of 
($i-$[4.6])$>4$  or ($i-K$)$>3.6$ is nearly as clean.  
RIGHT: Same plot but with  ($i$ - $K$) reveals no clear advantage to
use of the WISE mid-IR magnitudes.  Large red diamonds show colors
from the stellar SED compilation of \citet{Kraus07}.
  \label{imW2_gmr}}
\end{figure}

The red end of the FHLC sequence is also not inconsistent with the MTB
scenario; dC stars taper off at colors equivalent to mid-M dwarfs.
This is also where the SDSS magnitude limits probe too low a volume to
have significant surface density of later types (see e.g.,
Figure\,\ref{imW2_gmr}). 


\section{Multi-wavelength Data}  \label{sec:multiband}

Since many of our C stars have significant proper motions,
we have to be somewhat liberal in the positional search radii used for
matching to multiwavelength catalogs from imaging obtained
at a variety of epochs. Unless otherwise specified,
we  retrieved and matched catalogs from the Virtual Observatory (VO)
using the TOPCAT tool.  

We matched all objects in our SDSS C star catalog to the 2MASS
All Sky catalog using a 3\arcsec\, search radius.  99\% of matches
found were within 2\arcsec.   We also matched the C star catalog to
the UKIDSS DR6 to within $r=$2\arcsec, but found no matches at
$r>1.5\arcsec$.  

The Galaxy Evolution Explorer (GALEX) all-sky UV survey
\citep{Martin05} makes it feasible to detect hot white dwarfs in
unresolved binaries with main sequence companions as early as G-type
and K-type, and cooler WD with companions early M-type or later.  We
matched to GALEX GR6 UV catalog using a 4\arcsec\, search radius, but
found no matches beyond 2\arcsec.  Our shift and rematch experiments
yield a $\sim$3\% spurious match rate.  GALEX GR6 covers most of the
SDSS footprint.

To search for mid-infrared (3.4, 4.6, 12, and 22$\mu m$) photometry,
we uploaded our object catalog to
IRSA\footnote{http://irsa.ipac.caltech.edu/cgi-bin/Gator/nph-dd} for
matching to the Preliminary Release Source Catalog from NASA's
Wide-field Infrared Survey Explorer (WISE; \citealt{Wright10}) using a
4\arcsec\, search radius.  90\% of matched WISE sources have total
WISE astrometric uncertainty $<1.1\arcsec$ (0.8\arcsec\, per
coordinate).  Actual SDSS/WISE match separations cutoff sharply at
about 2\arcsec, with a tail (encompassing some 14\% of objects) out
to 4\arcsec.

\subsection{GALEX Ultraviolet-Detected Stars}  \label{GALEXdets}

 
GALEX photometry is in two bands, far- and near-ultraviolet
($F$ and $N$) with $\lambda_{\rm eff}=$ 1516\AA\, and 2267\AA, respectively.
Strong UV flux such as might be detected by GALEX in stellar
systems can arise from a very hot blackbody,
such as a hot white dwarf, or may also be associated with
stellar activity \citep{Pagano09}.  In our full sample of 1390 SDSS
stars showing carbon (including DQ WDs), we find 318 GALEX detections.    

We may expect that FHLCs with GALEX detections should arise primarily
from systems with hot white dwarf companions.  As evidence for our
capacity to detect such systems in GALEX GR6, 136 of the 318 GALEX
detections are DQ white dwarfs, which is 80\% of the (167) DQs in
our sample.  Excluding the DQs leaves 182 of 1220 FHLCs detected
by GALEX, or 15\%.  This is about 5 times the (3\%) spurious match
rate, so the majority are truly associated detections.

\begin{figure}
\plotone{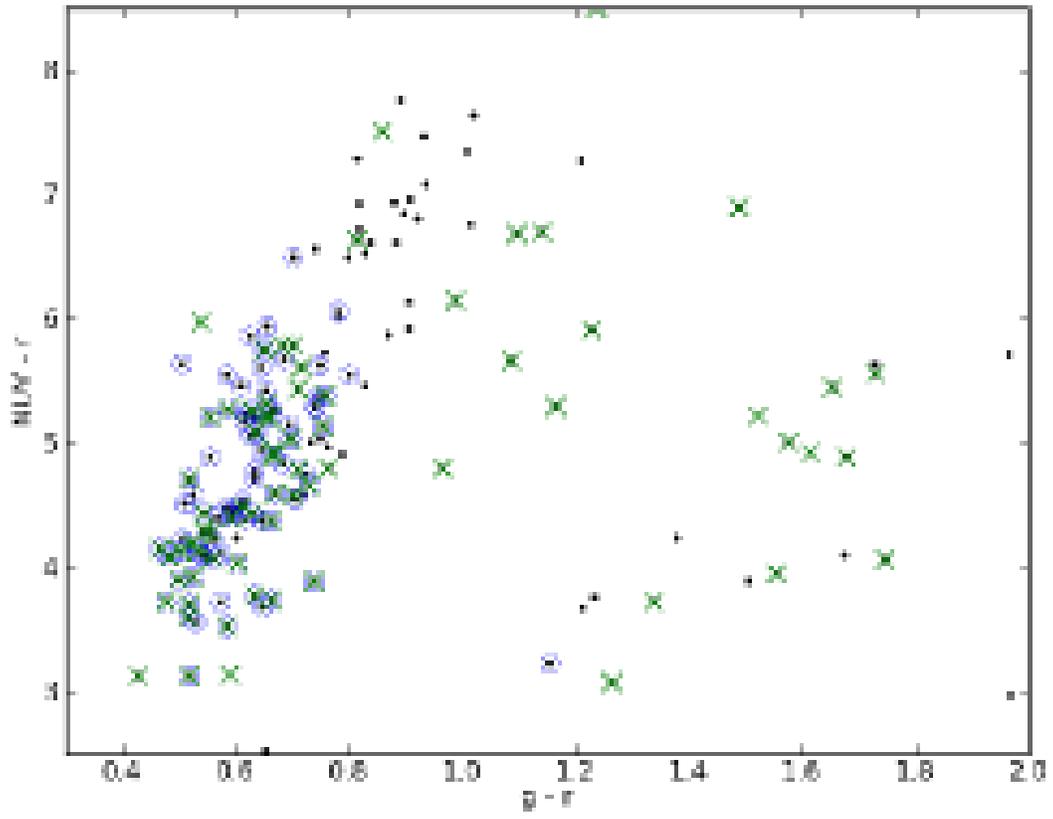}
\caption{Plot of (NUV-$r$) vs \gmr\, colors for SDSS stars in our sample 
(black dots).  G-type FHLC stars are shown as open blue circles.  
Green crosses indicate objects with significant proper motion.
  \label{gmr_Nmr}}
\end{figure}

Of these 182 GALEX-detected FHLCs, 90 are G types.  These 90
represent 68\% of all the (134) G-type FHLCs in our sample, an
extremely high detection fraction.  In comparison, the fraction for
``typical'' FHLC stars (neither G-type nor composite) is only 8\%.
However, G stars are both brighter and bluer than typical FHLC
stars.  Figure\,\ref{gmr_Nmr} shows the  (NUV-$r$) vs \gmr\, 
colors of our FHLC sample. 
Limiting both G and typical samples to be brighter than the
75th percentile of $r=18.27$, the detection fractions are 76\% and
21\%, respectively.  Limiting instead both samples to be bluer than
the 75th percentile of \gmr =0.666, the detection fractions are
70\% and 45\%, respectively.  We therefore conclude the G-type FHLC
stars indeed have stronger NUV fluxes than other FHLC stars, independent 
of their optical colors.  

However, when we match the predominantly normal (high proper motion)
stars in the \citet{Gould04} catalog to GALEX, we find that stars
with G-type colors \gmr$<0.8$ have a 72\% GALEX detection rate,
compared to a 6.5\% rate for redder stars.  In other words, a GALEX
NUV detection pulls out the bluer brighter stars (most likely
the metal-poor turnoff stars) at a rate approximately independent of
carbon enhancement. So, we conclude that the high NUV detection rate
of G-type FHLC stars is not evidence for hot white dwarf companions.

Only 3 of the 90 NUV-detected G-type FHLCs have FUV detections,
which could signal hot WD companions. Besides having a hot white 
dwarf component, UV brightness can arise in
young, active objects, from their active regions, transition regions,
or chromospheric emission.  However, none of the G-type FHLC stars
show emission lines. 

After excluding the G-types, of the remaining GALEX-detected FHLCs,
10 indeed have strong emission lines.  One of those is Draco C-1, a  
known symbiotic star in the Draco dwarf galaxy.  
Of the 9 galactic GALEX-detected objects with emission lines, 6 are
clearly DA/dC composites, showing evident broad Balmer absorption in
the blue.   

Indeed, of all 17 non-G-type galactic objects detected in the FUV
band, 7 are composite DA/dC systems.   
It is of interest to determine what sorts of non-G-type,
non composite FHLC stars are FUV-detected.  Of 9 such systems
two are clearly giants, with $r<16$ and undetected proper motion.
SDSS\,J 063313.18+840900.1 and 144945.37+012656.1 probably
have hot white dwarf companions that are overwhelmed in the 
optical by the luminous cool giant.  At least one or two others
may be G-type or composites misclassified due to low S/N spectra.
We venture that the rest, especially those with insignificant
proper motions may be subgiants harboring hot white dwarfs,
or possibly spurious matches near QSOs.

Red giants are cool, so should not have significant UV emission unless
a WD companion is accreting material thrown off by the RG wind, i.e.,
a symbiotic star.  In such a case, very soft X-ray emission may be
detected as well, and strong emission lines should be present. As
an example, Draco C-1 is detected by both ROSAT and GALEX and has
strong high ionization optical emission lines; \S\,\ref{sec:rosat}.

We consider that cool (non G-type, i.e., $\gmr>0.8$)
FUV-detected FHLC stars without detectable proper motions 
or strong emission lines are quite likely to be dC systems.

\subsection{Near Infrared Properties}   \label{sec:NIRprops}

We match the sample to the 2MASS point source catalog.
\citet{Joyce98} explained the unusual location of dC stars in the
JHK color-color diagram, via collision-induced absorption, which is
predicted to suppress molecular absorption features in the
near-infrared for stars of low metallicity and high surface gravity.
\citet{Downes04} noted that a large fraction of objects 
satisfying a more restricted color range of 
$(J-H)<1.2$ and $(H-K)>0.29167\,(J-H) + 0.25$ were dwarfs.

\begin{figure}
\plottwo{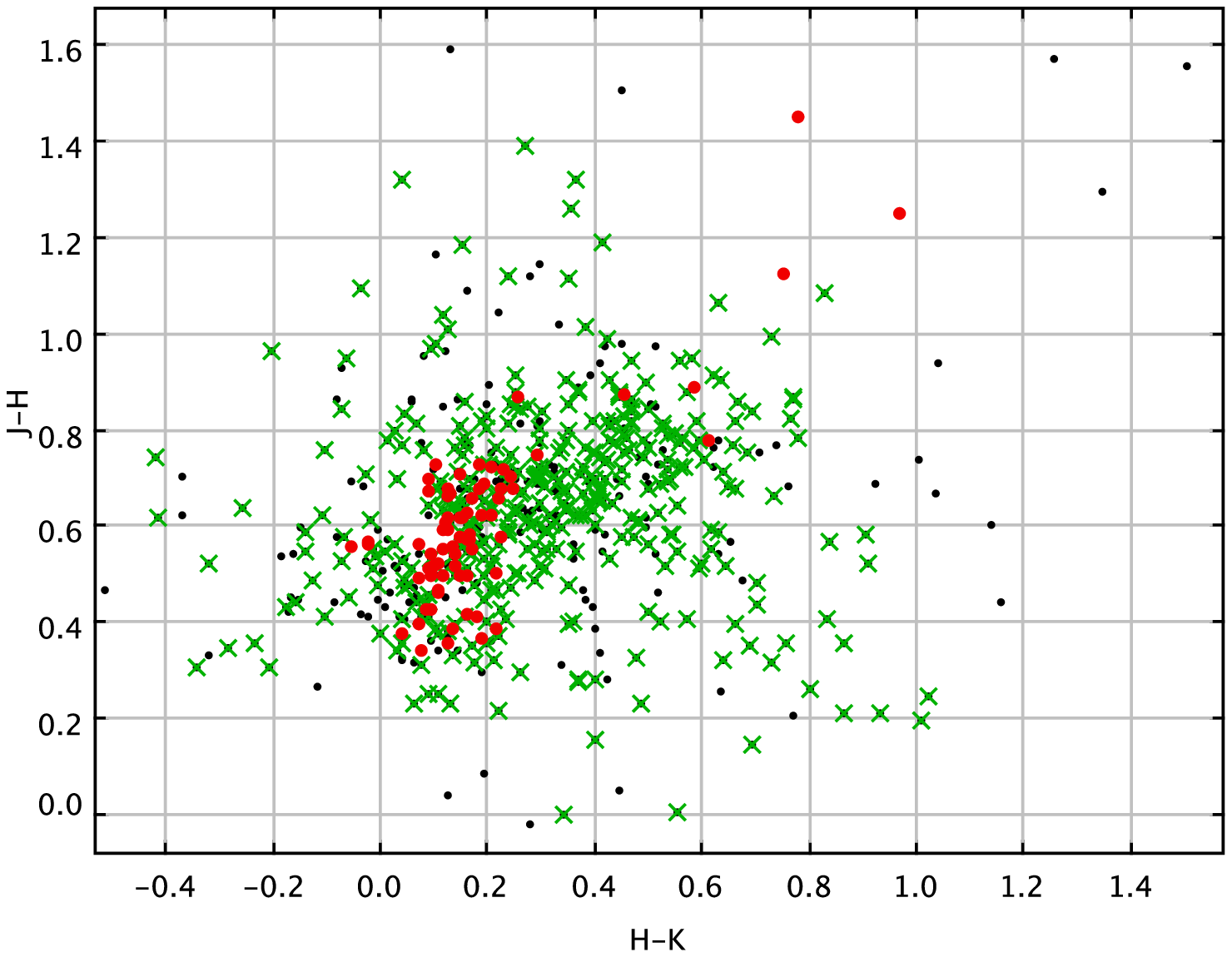}{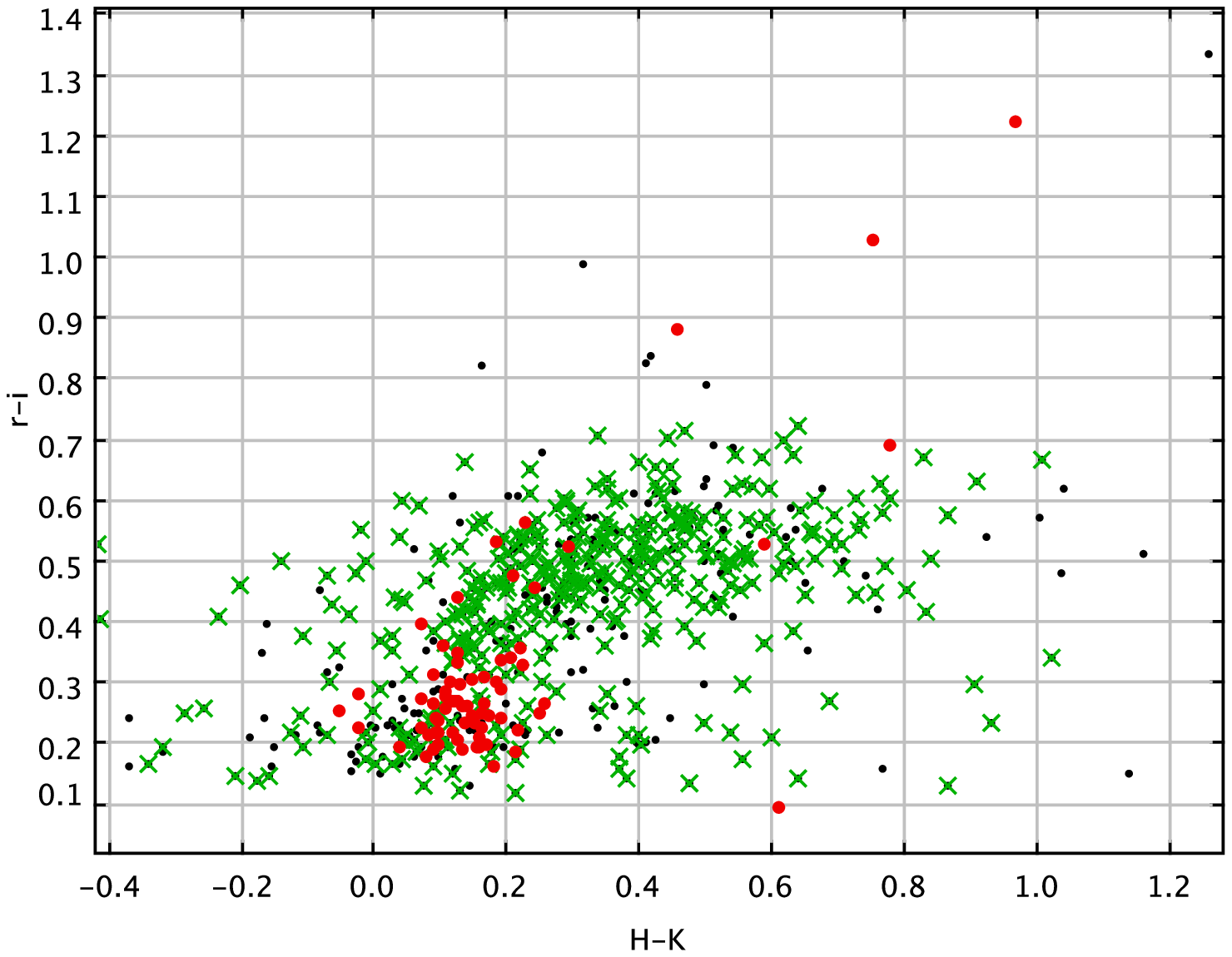}
\caption{Plots of (J-H) vs. (H-K) colors by luminosity class,
as determined in \S\,\ref{sec:absmags}.
LEFT: Within the magnitude range of our (SDSS spectroscopic) sample
of FHLC stars, near-IR colors seem to distinguish a region 
$(H-K)>0.3$ where the vast majority of stars are dwarfs (green
crosses), with most giants (red dots) clustered at bluer values.
RIGHT: Use of \rmi\, color for the ordinate axiscreates an even
stronger segregation of giants within our sample.  However, see
Figure\,\ref{HmK_JmH}.
 \label{IRlums}}
\end{figure}

\begin{figure}
\plotone{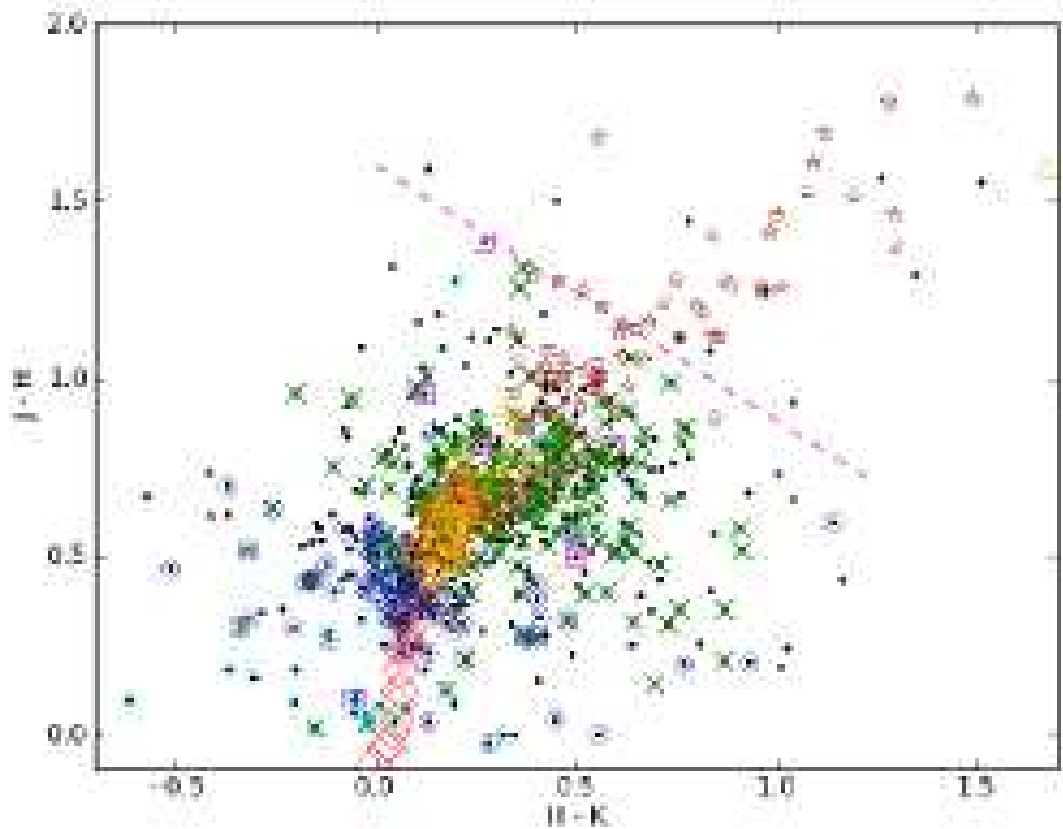}
\caption{Plots of (J-H) vs. (H-K) colors for FHLC stars with UKIDSS or
  2MASS detections.  Incorporating a wide variety
of C stars, we include our SDSS FHLC sample, SDSS DQ white
dwarfs (open cyan boxes), as well as N stars (open red stars) and R stars
(open orange triangles) from  \citet{Alksnis01}.  Essentially no star
redder than the line  $\jmh=-0.72\hmk\,+$1.6 shows a significant
proper motion.   
  \label{HmK_JmH}}
\end{figure}

Figure\,\ref{IRlums} shows that 
within the magnitude range of our SDSS spectroscopic sample
of FHLC stars, near-IR colors seem to distinguish a region 
$(H-K)>0.3$ where the vast majority of stars are dwarfs (green
crosses), with most giants (red dots) clustered at bluer values.
Use of \rmi\, color for the ordinate axis creates an even
stronger segregation of giants, at least within our sample.  However,
we see that in Figure\,\ref{HmK_JmH} inclusion of a wider sample of
known C giants further weakens the apparent strength
of near-IR colors alone as a photometric luminosity discriminant. 


However, cool N-type AGB C stars tend to populate the region
$(J-K)_0 > 1.4$ and $(H-K)_0> 0.45$, a selection used
by \citet{Demers07} to select thin/thick disk velocity probes.
We find 37 stars that satisfy these criteria. 


\subsection{WISE Mid-Infrared-Detected Stars}  \label{sec:WISEdets}


Since dCs are thought to have accreted mass dumped from an AGB star,
there may be cases where a redsidual debris disk can be detected.

WISE surveyed the entire sky at 3.4, 4.6, 12 and 22 microns in 2010,
achieving 5-$\sigma$ point source sensitivities better than 0.08,
0.11, 1 and 6 mJy in the four bands (for unconfused regions on the
ecliptic plane). The WISE All-Sky data release (March 14, 2012)
includes 563,921,584 point-like and resolved objects with S/N$>5$
in at least one band.  Of the 1035 FHLCs matched
to SDSS, we eliminate from our analyses the small
fraction (2\%) of sources that have any saturated pixels or 
are deblended ({\tt nb}$=1$), or have more than one PSF component in
the profile fitting ({\tt nb}$>1$).  

Figure\,\ref{imW2_gmr} shows optical vs. WISE colors for FHLC stars.
N-type AGB stars can be effectively separated in this plot by their
extreme redness; most FHLC stars with ($i-$[4.6])$>4$ have no proper
motions, and are likely on the AGB.  A similar plot using ($i-K$) is
equally effective.  FHLC stars in these plots appear to be offset
from the main sequence of high-p.m. dwarfs merely by their redder
\gmr\, colors, caused by \ctwo\, bands falling within the $g$ filter
bandpass (Figures\,\ref{dCspec} and \ref{Gtypespec}).

\section{Proper Motions}   \label{sec:pm}

The proper motion catalog of \citet{Munn04} combines astrometry from
the USNO-B and SDSS imaging surveys.  Our criteria for a significant
proper motion are as follows: (1) at least one USNO-B detection and
one SDSS detection per source (nfit $>$ 2), (2) proper motion in at
least one coordinate larger than $3~\sigma$, where $\sigma$ is the
proper motion uncertainty in that coordinate, and (3) total
proper motion larger than 11~mas\,yr$^{-1}$.

Reduced proper motion (RPM) correlates well with absolute magnitude  
\citep{Stromberg39}.  Where the absolute magnitude $M=m + 5 +
5\,log\,\pi $ (where $\pi$ is the parallax in arcsec), RPM
are calculated as $m + 5\,log\,\mu$ (where $\mu$ is
the total proper motion in arcsec per year).  If all stars had
te same transverse velocity, only a zero-point shift would be needed
to match a color-RPM to a color-magnitude locus. In an color-RPM diagram,
subdwarfs should be offset from main sequence stars because they have
large transverse velocities and because at the same color they are
dimmer. 

\begin{figure}
\plotone{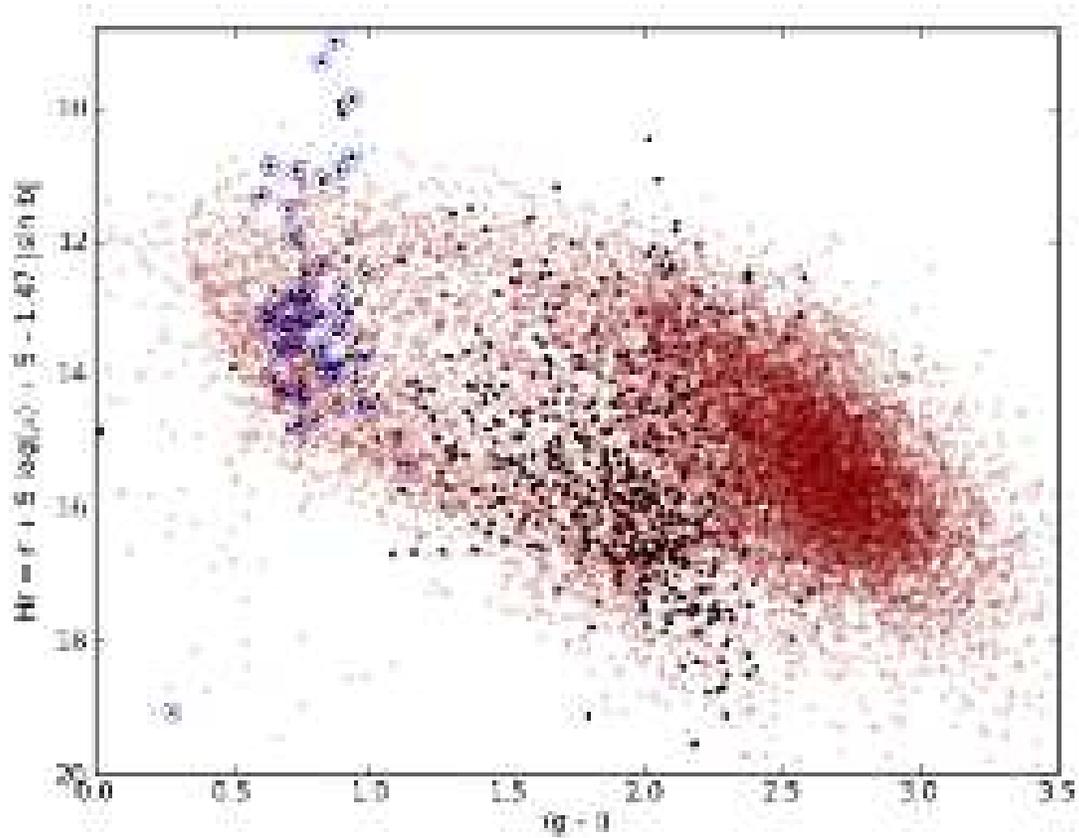}
\caption{Plot of the reduced proper motion vs. ($g-i$) color
for stars with significant proper motions.  Small red circles are
generic high proper motion dwarfs ($\mu>50$\,mas/yr) from the SDSS
DR1/USNO-B catalog of \citet{Gould04}.  G-type stars (blue circles)
appear to be blue, and may represent more massive subdwarfs.  Other dC
stars (black dots), appear to fall between the red dwarfs (densest
region on the right), and the subdwarfs (lower diagonal track to the
left), indicating they  may either be slower moving halo dwarfs, or
perhaps thick disk stars.  
  \label{gmiHr}}
\end{figure}

In Figure\,\ref{gmiHr}, we plot the reduced proper motion (RPM)
  $$H_r = r + 5\,{\rm log}\frac{\mu}{\rm
  mas\,yr^{-1}}\,-\,10\,-\,1.47\,|{\rm sin}\,b|$$
against $\gmi$\, color for stars with significant proper motions.  
The final term in $H_r$, dependent on Galactic latitude $b$,
compensates for the vertical offset of the RPM distribution of
samples of stars at different $b$ \citep{Salim03}.  

 The dC stars (black dots), appear to fall
below the red main sequence dwarfs (densest region on the right),
closer to the subdwarfs (lower diagonal track to the left), indicating
they may either be slower moving halo dwarfs, or perhaps thick disk
stars.   However, the apparent dCs' offset in $H_r$ is likely 
{\em weakened} by the fact that dCs of a given mass or absolute
magnitude appear redder than their O-rich main sequence counterparts
because of the strong \ctwo\, bands falling in the $g$ passband.
So we suggest that the true dC offset in absolute magnitude may indeed 
correspond to the subdwarf population.  G-type stars (blue circles) simply
represent earlier type, more massive subdwarfs.


\section{Absolute Magnitudes}   \label{sec:absmags}

The absolute magnitudes of carbon dwarfs are poorly known.
As part of the USNO parallax program, \citet{Harris98} found that
the 3 dCs they measured had very similar luminosities (9.6$<M_V<$10.0),
$\sim$2\,mag subluminous compared to normal disk dwarfs of similar
colors and solar-like metallicities.  This could be due
to low metallicity, enhanced He, and/or the strong molecular lines in 
the optical, but the kinematics are consistent with a spheroid
population.  

We find 7 dCs in our sample with additional ground-based parallax
measurements provided by Hugh Harris (private communication).  
For these, we find mean absolute magnitudes $M_r=9.0\pm 1.1$, 
$M_i=8.5\pm 1.1$, and $M_K=6.0\pm 0.9$.  The mean $M_V$
is $9.4\pm1.5$.  Within the small sample of these dCs with parallax
measurements, there is no significant correlation between absolute mag
and color, to encourage us to predict dC luminosity based on color. 
However, this is at least in part because the dCs with parallax
measurements span a relatively narrow color range of 
0.4$<$\rmi$<$0.6, with mean 0.49$\pm$0.07.  
If we exclude composite systems and G-types from our dC subsample
(yielding 656 objects), the mean \rmi\, is similar: 0.45$\pm$0.12.   
However, we certainly expect a range of absolute magnitudes for FHLC
stars, and in particular the G-types extend to \rmi$\sim$0.1, and show
smaller proper motions, and so are likely to be more luminous.



We examined a variety of color-color plots for the C stars, to see if
we could determine a color that closely tracks the O-rich main
sequence.  It is clear from the C star spectra that molecular band
features spanning a huge range of equivalent width routinely affect
the $g$ bandpass (spanning $\sim$4100 -- 5400\AA).  The $r$ band
($\sim$5600 -- 6800\AA) may also be affected, but less so, and the $i$
band even less ($\sim$6900 -- 8100\AA), except where strong CN bands
may occur.

\begin{figure}
 \includegraphics[scale=0.65,angle=0]{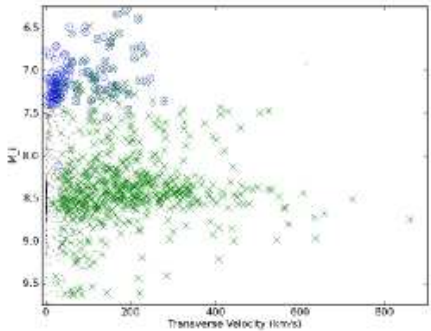}
\caption{Estimated absolute magnitude $M_i$ of FHLC stars
vs. transverse velocity $v_{\rm t}$, using techniques described
in the text.  Symbols are as in previous figures.  Since no G-type dCs
(blue circles) appear to be more luminous than $M_i\sim 6.3$, we adopt
this as our criterion to distinguish C giants from dCs. 
\label{vtMi}}
\end{figure}

We had hoped that the near-IR colors would offer a smaller dispersion
around the O-rich main sequence, but the 2MASS-matched sample
shows wide scatter in all the near-IR colors.  C star colors appear
to hew closest to the O-rich main sequence locus in the \rmi\,
vs. \imz\, plot for the range 0.1$<$\rmi$<$0.8, which corresponds
to their spectral types $\sim$G1 through M1 and includes 96\% of
the C stars.  Since the locus of median colors in this color range
closely follows \imz\,$=0.6\,\rmi ~-~ 0.04$, either color alone 
could be used just as well, so we choose \rmi\, colors (since errors
are generally lower than for \imz).  For each star in our sample,
we can derive an estimate of the absolute $i$ magnitude $M_i$, and
using the main sequence absolute magnitudes from \citet{Kraus07},
despite the fact that their sequence consists primarily of disk
stars.  Using this method, however, we find that the derived $M_i$
values for dCs are $\sim$1.75\,mag brighter than those of
the parallax-measured dCs in the same color range.  For definite dCs,
we therefore derive $M_i$ values from the \rmi\,-$M_i$  relation of
\citet{Kraus07}, but adding an offset of 1.75\,mag to estimate their
true absolute magnitudes.  This admittedly crude approximation can
then be used to estimate a photometric parallax distance.  Further
support for our adopted method comes from the fact that transverse
velocities then taper off strongly by about 400\kms\, (rather than
$\sim800$\kms) with only a handful above 600\kms (Figure\,\ref{vtMi}).


\subsection{Definite Dwarfs or Giants}  \label{sec:dwarfORgiant}

Given the sensitive proper motion detection limits, does a significant
proper motion guarantee that the star is a dwarf?  Given an observed
proper motion, simply assuming that the star is gravitationally bound
to the Galaxy sets an upper limit on its distance.  The upper limit,
combined with the observed magnitude, sets a lower limit to
absolute magnitude.

\begin{figure}
 \includegraphics[scale=0.65,angle=0]{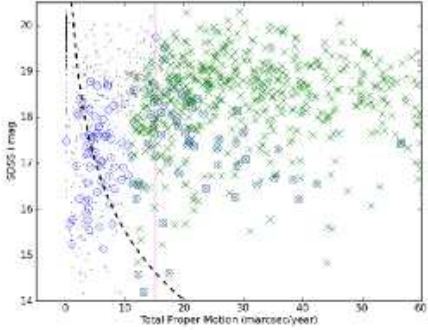}
\caption{Plot of SDSS $i$ magnitude vs. total proper motion
in mas/year for FHLC stars.  Stars with green crosses have
significant proper motions, using the criteria in 
\S\,\ref{sec:pm}.  The vertical magenta line shows
our 90\% proper motion detection limit of 15\,mas/yr.  Blue circled
points are G-type FHLCs.   If we assume that the Galactic escape
velocity is 600\,km/sec, the green dashed line shows for each proper
motion the magnitude fainter than which the star must be a dwarf ($M_i>3$).
Essentially all FHLC stars with detected proper motions in our sample
are therefore dCs, with the possible exception of a handful of bright
G-type FHLC stars, which may be subgiants.
\label{totPMimag}}
\end{figure}

Recent estimates from the RAVE survey \citep{Smith07} place the local
Galactic escape velocity lies within the range 498$<v_{esc}<608$km/s.  
We conservatively assume that the Galactic escape velocity $v_{esc} =
600$\kms.  Subgiant stars can be found at absolute $i$ band magnitudes
$M_i\lax 3$.  We adopt that anything less luminous than $M_i=3$ must a
dwarf. Figure\,\ref{totPMimag} shows that indeed virtually every object
with detected proper motions is definitively a dwarf.  A handful
of possible exceptions that tiptoe beyond the subgiant boundary
are G-type FHLC stars, which may be expected as they are
near the main sequence turnoff.  We adopt that any FHLC star in our
sample with a detected proper motion as defined in \S\,\ref{sec:pm} is
a dC, and there are 729 examples, marked Class=d in
Table\,\ref{tab:fhlc}, about 6 times as many as previously known
\citep{Downes04}, and 59.7\% of our total FHLC sample.

%

Conversely, can we determine that an FHLC star is definitively a
giant?  We follow the line of argument established in
\citet{Downes04}.  First, we estimate our detection limit for proper
motions, which we determine to be $\mu^{\rm min}\sim 15$\kms\, because 
90\% of objects with significant proper motion exceed that value.
Then, using the $M_i$ estimates established above,
we derive a photometric parallax distance for objects with a
significant proper motion, and use that to calculate transverse
velocity $v_{\rm t}$.  A cumulative histogram reveals that
90\% of dCs have $v_{\rm t}>50$\kms.  Assuming that dCs have 
$v_{\rm t,min}=50$\kms, then a non-detection places a  
{\em lower} distance limit on an FHLC of 
    $$d_{\rm min}= v_{\rm t,min}\,(1000/4.74\,\mu_{\rm min}) =
1055\,{\rm pc}$$
where $v_{\rm t,min}$ is in km/s and $\mu_{\rm min}$ is the total
proper motion in mas/yr.  From $d_{\rm min}$, we find the
least luminous absolute magnitude $M_{\rm i,max}$ that a
$v_{\rm t}\sim50$ star must have to have escaped proper motion
detection, and compare that to a dwarf/giant threshold criterion.
Figure\,\ref{vtMi} shows that dCs quite likely achieve magnitudes as
luminous as $M_i^{\prime}\sim 6.3$, which we therefore adopt as our
threshold.  Any FHLC star with a measured but insignificant proper motion, but
an $i$-band magnitude 
       $$i < M_i^{\prime} -  5 + 5\,{\rm log}(d_{\rm min}) = 16.4$$
we consider to be a definite giant, and are marked as such
(Class=g) in Table\,\ref{tab:fhlc}.  With this definition, we find
59 giants, or 4.5\% of our full FHLC sample.

Objects with no proper motion measurements, or insignificant measured
proper motions and $i>16.4$, are marked as Unknown (Class=u). 
There are 419 objects in this class, or 34.3\% of our sample.

If we restrict consideration to the 1050 objects with existing proper
motion measurements, the fraction with dwarf, giant, or unknown 
luminosity class becomes 69.4\%, 23.7\%, and 6.9\% respectively.


\section{Velocities}   \label{sec:rvs}

\begin{figure}
\plottwo{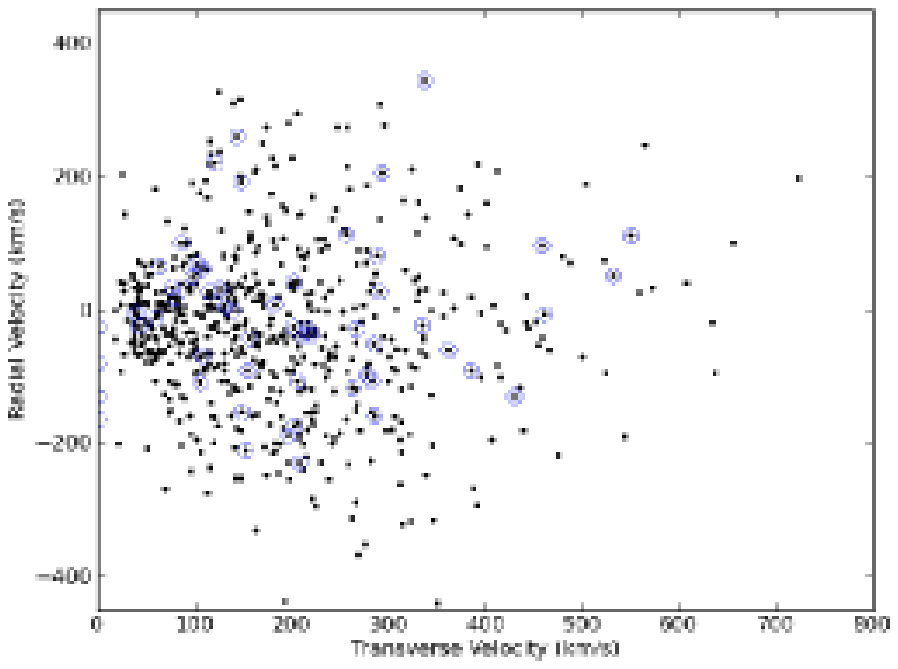}{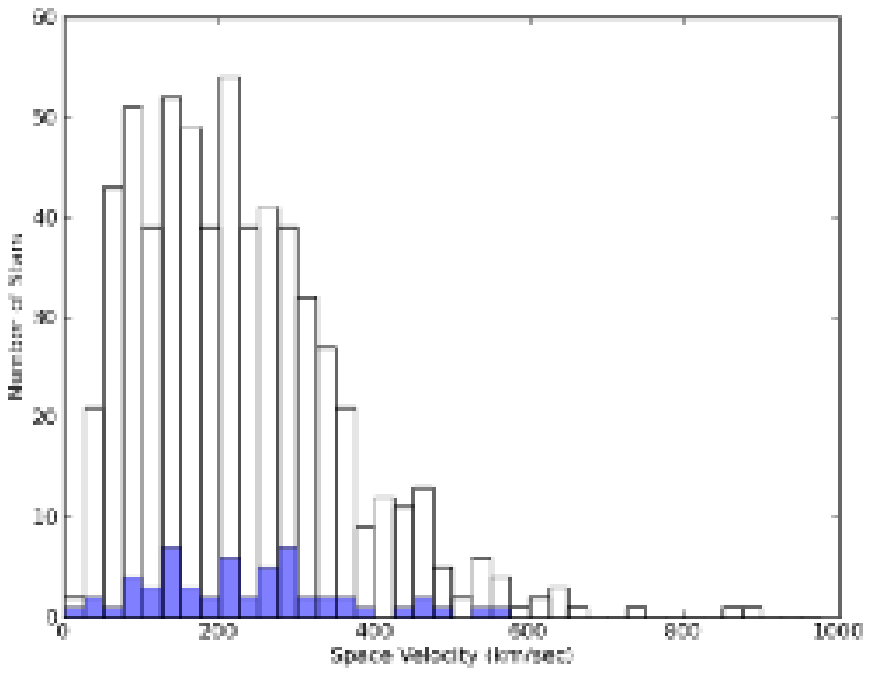}
\caption{LEFT: Plot of the radial velocity derived from comparison to the
ELODIE templates vs. estimated transverse velocity for 621 FHLC stars
with reliable proper motions  (see \S\,\ref{sec:pm}) and colors
suitable for absolute magnitude estimates (see \S\,\ref{sec:absmags}).
G-type FHLC stars are shown as blue circles.
RIGHT: Histogram of space velocities $v_{sp}$ computed for the same
sample. G-type FHLC stars are represented by the shaded blue
histogram. 
  \label{vtrv}}
\end{figure}

SDSS tabulates radial velocities obtained two ways: from
cross-correlation using SDSS commissioning spectral templates
\citep{Stoughton02}, and from comparing to the (resolution-degraded)
ELODIE library spectra  as described by \citet{Moultaka04}. While for
typical stars, the ELODIE velocities are considered more reliable, the
ELODIE library does not contain any carbon stars.  Even for CEMP
stars, which lack the striking  \ctwo\, and CN bandheads of our
FHLC stars, \citet{Lee08} noted that cross-correlation with special
templates based on synthetic spectra are generally required to produce
reliable radial velocities.  Therefore, we judge a detailed study of
the kinematical properties of the population to exceed the scope of
this paper.  We nevertheless show in Figure\,\ref{vtrv} the
radial velocities of the sample plotted against derived transverse
velocities $v_t$. Since $v_t$ estimates require significant proper
motions (\S\,\ref{sec:pm}), as well as distance estimates based on our 
estimates of absolute magnitude, as described in \S\,{sec:absmag},
the plotted sample is limited to the 621 FHLC stars fulfilling those
criteria. The strong overlap of the redder dCs with the G-types,
illustrate that they likely originate in a similar population.
Figure\,\ref{vtrv} (right panel) shows a histogram of the
estimated space velocities $v_{sp}$ of the same sample.
The median space velocity is 208\,km/s, while the mean is
225$\pm$133\,\kms.  The distribution of $v_{sp}$ for the
G-type dCs in this sample is not significantly different than
the overall distribution.




\section{``N''-type AGB stars}   \label{Nstars}

The classical ``N''-type AGB C stars have very red colors and
often show strong CN bands, in some cases with Balmer emission lines
as well. Such stars' spectra are quite distinct in appearance from
dCs, unlike the spectra of warmer R or CH giants.  As such, faint
examples may be useful tracers of the Milky Way gravitational
potential at large Galactocentric distances.

We searched our sample for spectra noted as being ``red'' and/or with
``strong CN'', and removed objects with weak \ctwo\, lines or (in just
one case) a detectable proper motion.  This left 14 stars, 2 of which are
known in the dSph galaxy Leo\,II.  One other star,
SDSS\,J125149.87+013001.8, was recognized as an N star in
\citet{Downes04}.  The remaining 11 are all likely newly discovered
high galactic latitude N stars. 

The 17 Galactic N stars measured by Hipparcos have mean absolute
magnitudes $M_V=-0.8\pm 1.3$ and $M_K=-7.0\pm 1.1$ (with similar
median values; \citealt{Wall98}), which we take as $-1$ and
$-7$ respectively, given how uncertain we are about the similarity of
that sample to our own.

Using $M_K=-7$, the N stars in Leo\,II yield photometric 
distances of 235\,kpc for \\
SDSS\,J111312.83+221114.1 and 200\,kpc for
SDSS\,J111320.64+22 1116.3, consistent with published distances
\citep{vdB00}.  The remaining N star candidates range from 30 to 160
kpc in ($K$-band estimated) distance.

We note that the two N star candidates SDSS\,J144448.86-011056.2 and
SDSS\,J144631.08-005500.4 are within 30\arcmin\, of each other, and
having similar magnitudes, colors, and velocities, may well signal
the existence of a dwarf galaxy or tidal stream.  Their $K$ magnitude
($\sim$11) places them at a distance from us of about 40\,kpc, with
mean Galactic coordinates $l\sim 352\deg$, $b\sim 51\deg$. No
overdensity in this region is previously noted in the catalog of
neighboring galaxies of \citet{Karachentsev04}, but we suggest that
deeper images and color-magnitude searches may reveal something in the
vicinity. 

We note that {\em every} star that met our N star candidate criteria
has $(J-H)>$0.8, as do all the bright N stars from Alksnis' (2001)
catalog shown in Figure\,\ref{HmK_JmH}.  None of our N star candidates has a
GALEX detection.  Therefore, we are confident that, while it may not
be complete, our method is pure, i.e., it does produce a list of 
{\em bona fide} AGB N stars.

%

\section{C Stars in Local Group Galaxies}  \label{sec:localgroup}

We assembled a list of Local Group dwarf galaxies from the
\citet{McCon12},\footnote{Also thanks to Richard Powell for compiling
http://www.atlasoftheuniverse.com/galaxies.html.}
 having distance moduli $<$24.5\,mag.  At these distances,
the most luminous AGB C stars would have $V\lax$20.5, and therefore
could be bright enough for SDSS fiber spectroscopy\footnote{80\% of
our FHLC stars have $V\lax20.5$.}.  We find 10 FHLC stars from our
catalog that fall within a degree of their approximate central 
Local Group galaxy coordinates.  Two of these have significant proper
motions, and are likely to be foreground dC stars.  


Of the remaining 8, 3 are known to be in Draco\footnote{In fact, 
  SDSS\,J171942.39+575837.7 has no proper motion measurement, but has
  $v_r=-278\pm3$, quite similar to the other known Draco C
  stars.}\citep{Downes04}.    

We also find 3 C stars already known in Leo~II \citep{Aaronson83,Vogt95}.   
At the distance of Leo\,II ($\sim$205\,kpc), they have
$M_r\sim -2.8$. SDSS\,J111324.84+220916.8 has seemingly bizarre \gmr\,
colors and a significant proper motion measurement, but both are due to
blending with nearby objects in Draco.   

One C star, SDSS\,J132755.56+333521.7 was recently noticed by
\citet{Zucker06} among SDSS spectra of stars in a newly discovered
local group dwarf galaxy in Canes Venatici. It has $M_r\sim -2.5$,
typical of a CH carbon giant.  

The one remaining star SDSS\,J095955.57+304058.5 with $r=20.61$ may well 
a member of Leo~A.  There is no proper motion measurement available
for this object; a firm upper limit to its motion would strengthen the
argument for membership.  If this object is a luminous AGB star
($M_r\sim-3.6$) and, by comparison to the known Local Group C stars
herein has ($r-K$)$\gax$2, then at the distance of Leo~A ($\mu=24.2$),
the expected near-IR magnitude $K\sim 18.6$ is too faint for
2MASS, and indeed the object is not detected therein. 

Leo~A is a dwarf irregular (dIrr) galaxy. C stars are less common, but
not unknown in dIrr galaxies. For instance the Magellanic dIrr DDO 190
(also known as UGC 9240) contains many (photometrically-identified) C
stars, which is taken as evidence for the existence of an old stellar
halo \citep{Battinelli06}.  Spectroscopically-confirmed C stars in
dIrr galaxies appear to be rare in the literature, so this object may
seems likely to be a valuable addition.


\section{X-rays from Carbon Stars}   \label{sec:xrays}

Since we now know that most C stars are in mass transfer
binary systems, a fraction of them may still have hot WD companions. 
For close binaries, there may be accretion onto the WD, or even
coronal activity in the C star as a result of spin-up from accretion
and/or tidal locking.

We searched several X-ray catalogs for positional coincidence
to our sample, using a somewhat generous matching radius, permissible
because both X-ray sources and carbon stars are rare on the sky.
We searched the Chandra Source Catalog (version 1.1;
\citealt{Evans10}) with a 6\arcsec\, radius and found no matches.   
Our results from XMM-{\em Newton} and ROSAT catalog searches are
described below. 

\subsection{SDSS\,J125017.9+252427.6}  \label{sec:xmm}

We searched the 2XMMi-DR3 \citep{Watson09} with a 6\arcsec\, radius
and found a single match to SDSS\,J125017.9+252427.6.  
SDSS\,J125017.9+252427.6 is a bright ($r=16.4$) dC star with
spectacular bands of \ctwo\, and CN, along with Balmer emission lines
visible from H$\alpha$ to H$\epsilon$.  It has a highly significant
proper motion (34.4$\pm$4\,mas/yr).  The positional separation from
the \XMM\, source is just 1.8\arcsec, while the nearest $g<22$ optical
counterpart is at 35\arcsec.  Its optical and IR colors are not
unusual for a dC, and it is not detected by GALEX.  Assuming $M_r=9$
for a dC puts this object at $\sim$300\,pc.  From the XMM reported
flux (0.2-2\,keV) of 6.36$\times 10^{-15}$\fcgs, we derive an X-ray
luminosity log\,$L_X=6.9\times 10^{28}$\lcgs.  Such a luminosity
is rather typical for X-ray detected stars in the Chandra
serendipitous survey (CHESS; \citealt{Covey08}).


\subsection{ROSAT Detections}   \label{sec:rosat}

We also searched the ROSAT All-Sky Survey Faint Source Catalogue
(RASS-FSC; \citealt{Voges00}), using a 25\arcsec\, radius.
This radius yields a spurious matching rate of about 35\% for SDSS
magnitude limits \citep{Agueros09}.   

We rediscovered that the known C star in the Draco dwarf galaxy, Draco
C-1 \citep{Aaronson82} is a ROSAT detection.  Draco C-1 is also
detected by GALEX, and as expected, shows no significant proper
motion. Its SDSS spectrum features powerful H$\alpha$ emission
($W_{\lambda}=110\pm10\AA$) with a broad base, and strong emission
lines of He\,I and He\,II, as noted by \citet{Margon02}.  Draco C-1 is
a known symbiotic star (a red giant transferring mass to a WD
companion), and variable in both flux \citep{Kinemuchi08} and radial
velocity \citep{Olszewski95}, the latter probably caused by
orbital motion.

\section{Completeness Studies}   \label{sec:FAST}

Our SDSS sample therefore represents a heterogeneous sample, but
is not strongly biased toward the initial FHLC star selection color
wedge.  However, even within the SDSS color wedge (and certainly
without) the fraction of objects that were sampled, and which turn out
to be C stars is a complex function of competing SDSS fiber assignment
priorities and their evolution. We therefore have undertaken our own 
experiment specifically to constrain the fraction of dC stars within
the color wedge.


Using the 1.5m Tillinghast reflector and FAST spectrograph
at the Fred L. Whipple observatory on Mt Hopkins, we obtained
FLWO1.5m/FAST spectroscopy of dC candidates.  To arrive at our FAST
sample, we chose from SDSS DR7 the color locus where $>$90\% reside, which
corresponds closely to the color plane ($g-r$)$> 1.587$ ($r-i$) +
0.534 examined by (\citealt{Margon02}, see our Figure\,\ref{rmi_gmr}).
We then require a minimum proper motion of 11\,mas/yr from SDSS/USNOB,
combined with reasonable quality criteria (e.g., p.m.$>3\sigma$,
$r>$15 to avoid saturation). Among the resulting 171,650 objects,
there were 2334 with SDSS-II spectra.  Inspection of these yielded 165
{\em bona fide} dCs, 43 more more than published at the time.
However, all those were in a narrower wedge of color, allowing further
restriction by ($r-i$)$> 0.3556\,$($g-r$) - 0.17
(Figure\,\ref{rmi_gmr}).   


To construct a pilot spectroscopically complete sample, we selected a
random subsample of 300 stars 
within this narrower color wedge, also having high quality photometry, total
proper motion $>40$mas/yr and observable efficiently with FAST:
$15<r<17$ and $\delta>-05$.  Our pilot sample is uniform, homogeneous,
and randomized so that completeness is easily derived.  From May 2009
until May 2010, we obtained 291 FAST spectra, spanning 3470 --
7400\AA\, at 1.5\AA/pixel resolution.  We find 10 of these are {\em
  bona fide} new dC stars, listed in Table\,\ref{tab:FASTdcs},
which corresponds to a fraction (from the
$\beta$ function distribution) of $3.6\pm 0.1$\%.  Three of these
objects also have SDSS spectra.  Our sample of high quality dC spectra
are rather typical, except for one object discussed below.


Our final photometric criteria are shown explicitly as CasJobs query
language in Appendix\,\ref{sec:FASTfinalquery}.  The full sample of
SDSS DR7 objects meeting these criteria constitutes 1,973 stars.
Applying the dC 
fraction from our observed FAST sample implies that there are 71$\pm$2
such bright, high proper motion dCs.  The SDSS DR7 sky area with
$\delta>-05$deg represents about 10,770\,degsq, so that the surface
density of such objects is about one per 150\,\degsq.  This number
can be used to constrain the dC population model, and thereby
the history of AGB stars and binary mass transfer in the Galaxy. 
Such modeling is well beyond the scope of the current work, since
a full model of dC populations should include their likely
distribution of absolute magnitude, color, space density, kinematics,
and scale height \citep{deKool95}. Indeed, several distinct
populations are likely needed, halo, thick and perhaps even thin disk.    

\begin{figure}
\includegraphics[scale=0.7,angle=-90]{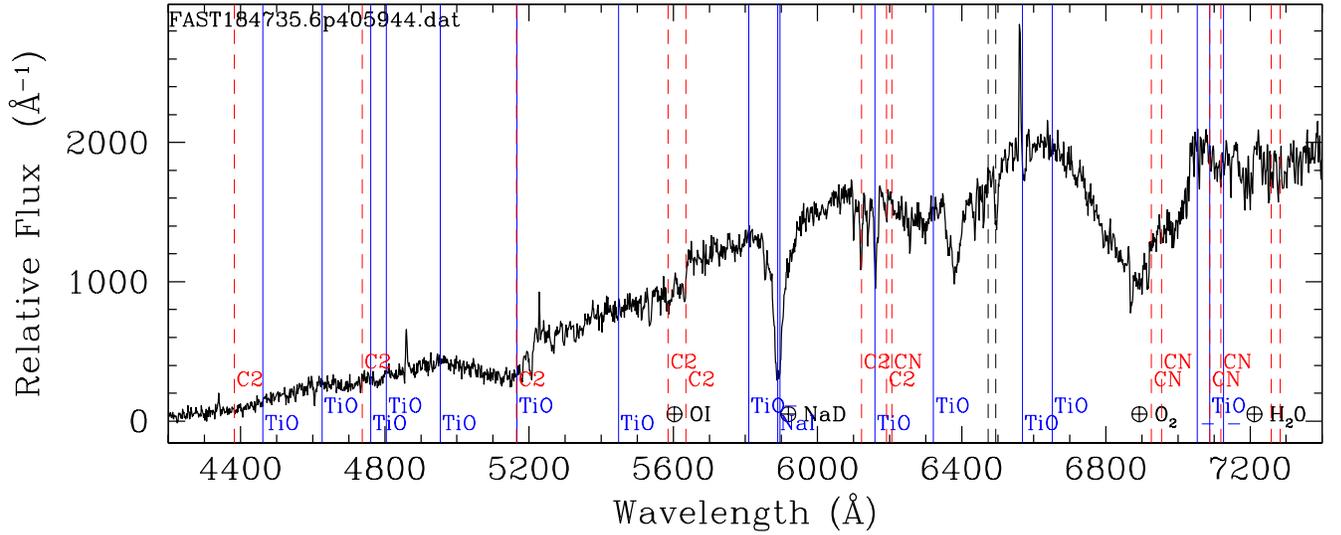}
\vspace{-5cm}
\caption{FLWO\,1.5m/FAST spectrum (flux in ADU) of the most unusual
of the 10 dC stars we discovered in our FAST completeness program,
SDSS\,J184735.6+405944. This object shows relatively weak \ctwo\,
absorption, strong NaD absorption, strong Balmer emission lines, and
broad deep CaH absorption features centered near 6385\AA and
6900\AA. CaH is only this strong is dwarfs, and was noted in 
the case of SDSS\,J153732.2+004343 as a possible luminosity
discriminant in low-resolution spectra such as these. 
The locations of C$_2$ bands strong in most dCs are marked
with vertical red dashed lines; only the 5636\AA\, bandhead is
clearly evident. TiO bandheads strong in later M dwarfs are marked
with vertical blue lines.  None are seen, so we estimate the dC has an
M0V companion.
  \label{FAST184735}}
\end{figure}

As part of our FAST program we discovered the only known dC/dM
spectroscopic binary star, which must also harbor a white dwarf.
The spectrum of SDSS\,J184735.67+405944.2 in Figure\,\ref{FAST184735}
has clear (relatively weak) \ctwo\,
absorption, strong Balmer emission lines, and a broad deep absorption
feature centered near 6900\AA, corresponding to the 6909\AA, 6946\AA\,
CaH bands. Since this feature is not often seen in dCs, and since
the \ctwo\, bands appear weak, we believe that this is a spectroscopic
composite system that also contains an M dwarf.  Together with the
as yet unseen white dwarf, this represents the first known dC triple 
system. For the M star to still have C/O$<$1, it is likely farther
from the former AGB star than the dC, so we speculate that the Balmer
emission lines come from interaction between the dC and the white
dwarf, which is nevertheless too cool to contribute significantly to
this spectrum (and undetected by GALEX).   


\section{Conclusions}   \label{sec:conclude}

All the data we have gathered and analyzed here remain consistent with
the hypothesis that all dCs were ``innocent bystanders'' to mass
accretion in binary systems.  The more massive initial primary star
evolved up the AGB, producing a carbon-enhanced stellar wind accreted
by its companion main-sequence star, which now bears the atmospheric
traces of carbon as a dC star.


Most FHLC stars coincide in color-magnitude space with the locus of
``thick-disk-like'' ([Fe/H] $= -0.7$) stars \citep{deJong10}.  The
brighter G-type FHLCs instead coincide with turn-off stars of
older, thick-disk or 'inner-halo-like'' stars of intermediate
metallicity  ([Fe/H]$ =-1.3$).  We find a high (68\%) UV detection rate
of G-type FHLC stars, but being similar to the detection rate of
generic high proper motion stars of similar color,
this is not evidence for hot white dwarf companions.

Thus, they 
are clearly in systems where the intial primary has already turned off
the main sequence, and indeed completed the AGB phase.  If the mass
ratio $q$ of binary systems peaks near unity \citep{Reggiani11}, then
most dCs should be close to the turnoff mass, especially if the mass
accretion is enhanced in higher mass bystanders.  However, above
a certain mass and effective temperature, the formation of \ctwo\,
molecules is suppressed, and at some point the \ctwo\, features by
which our FHLCs are identified diminish below detectability.
Higher mass objects might still be identified e.g., as carbon-enhanced metal
poor stars, but only with spectroscopy of significantly higher
resolution and S/N. 

We may reasonably hypothesize that dCs have or will evolve into Ba or
CH giants or subgiants, with a range of properties dependent on their
masses and enrichment histories.  Further study of the binarity,
masses, and/or kinematics of this sample would help to investigate
this hypothesis.  Post-MTB dwarfs more massive than the G-type FHLC
stars currently show no signs of \ctwo, but while they are therefore
not included in our sample, we hypothesize that they must exist, and
are best found through spectroscopy of higher resolution, or by far-UV
detection of hot white dwarf companions.

Only a handful of the ``smoking guns'' - systems where the AGB remnant
is easily discerned in the optical spectrum or by UV excess -
have been found.  In the current work, we expand  from 2 to 8
the number of such systems known, and we describe them fully in a
companion paper \citep{Green13}.


\acknowledgements 
PJG is indebted always to Bruce Margon and Scott Anderson
for their encouragement and support.  PJG thanks Hugh Harris,
Andrew West, Kevin Covey and John Bochanski for useful discussions.  

Funding for the SDSS and SDSS-II has been provided by the Alfred
P. Sloan Foundation, the Participating Institutions, the National
Science Foundation, the U.S. Department of Energy, the National
Aeronautics and Space Administration, the Japanese Monbukagakusho, the
Max Planck Society, and the Higher Education Funding Council for
England. The SDSS Web Site is http://www.sdss.org/. 

The SDSS is managed by the Astrophysical Research Consortium for
the Participating Institutions. The Participating Institutions are
the American Museum of Natural History, Astrophysical Institute
Potsdam, University of Basel, University of Cambridge, Case
Western Reserve University, University of Chicago, Drexel
University, Fermilab, the Institute for Advanced Study, the Japan
Participation Group, Johns Hopkins University, the Joint Institute
for Nuclear Astrophysics, the Kavli Institute for Particle
Astrophysics and Cosmology, the Korean Scientist Group, the
Chinese Academy of Sciences (LAMOST), Los Alamos National
Laboratory, the Max-Planck-Institute for Astronomy (MPIA), the
Max-Planck-Institute for Astrophysics (MPA), New Mexico State
University, Ohio State University, University of Pittsburgh,
University of Portsmouth, Princeton University, the United States
Naval Observatory, and the University of Washington. 

This publication makes use of data products from the Two Micron All
Sky Survey, which is a joint project of the University of
Massachusetts and the Infrared Processing and Analysis
Center/California Institute of Technology, funded by the National
Aeronautics and Space Administration and the National Science
Foundation. This research has made use of the NASA/ IPAC Infrared
Science Archive, which is operated by the Jet Propulsion Laboratory,
California Institute of Technology, under contract with the National
Aeronautics and Space Administration.  

This research has made use of data obtained from the Chandra Source
Catalog, provided by the Chandra X-ray Center (CXC) as part of the
Chandra Data Archive, and from the 2XMMi catalog in the Leicester
Database and Archive Service at the Department of Physics and
Astronomy, Leicester University, UK.
\dataset[ADS/Sa.CXO#CSC]{Chandra Source Catalog v1.1}

This publication makes use of data products from the Wide-field
Infrared Survey Explorer (WISE), which is a joint project of the University
of California, Los Angeles, and the Jet Propulsion
Laboratory/California Institute of Technology, funded by the National
Aeronautics and Space Administration. 



\begin{figure}
 \includegraphics[scale=0.65,angle=0]{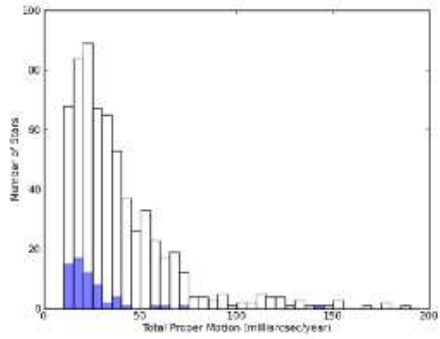}
\caption{Histogram of total proper motion for FHLC stars
with significant proper motions, using the criteria in
\S\,\ref{sec:pm}.  The blue shaded histogram shows the proper motion
distribution for the subset of G-type FHLCs.  The lowest bin is not
populated because of our criterion that the total proper motion be
larger than 11~mas\,yr$^{-1}$.  90\% of objects
with significant proper motion have total proper motion $>
15$~mas\,yr$^{-1}$, which we take as our detection limit.
\label{totpmHist}}
\end{figure}

\clearpage
\input{FHLC0}

\begin{deluxetable}{c}
\tablenum{2}
\tabletypesize{\scriptsize}
\tablecolumns{1}
\tablewidth{0pc}
\tablecaption{Stars in Downes et al. (2004) Not Included in Our Sample}
\tablehead{
\colhead{SDSS\.J Name}}
\startdata
015025.80-001315.1 \\
030837.07+005157.1 \\
032955.54-000354.0 \\
034705.41-063323.9 \\
074710.84+251619.6 \\
080046.72+364107.1 \\
090208.05+435503.8 \\
092545.46+424929.0 \\
100627.47+462117.4 \\
103758.57+612559.3 \\
104138.55+643219.8 \\
105629.95+012208.8 \\
111833.84+563958.1 \\
112836.57+011331.2 \\
113040.49+525039.2 \\
113931.62+050231.2 \\
115925.70-031452.0 \\
121733.35-001857.5 \\
121933.50-012843.0 \\
123834.40+023856.0 \\
143328.12+595808.9 \\
155043.84+571342.5 \\
165141.94+352012.6 \\
231340.54+143600.8 \\
\enddata
\label{Downes04NotHere}
\end{deluxetable}

\input{DQ0}
\input{FASTdc0}

\clearpage
\appendix

\section{SDSS Query for Final FAST dC Candidate
  Sample}  \label{sec:FASTfinalquery} 

{\scriptsize
\begin{verbatim}
-- SDSS DR7 CasJobs query to select carbon dwarf candidates in a color wedge
-- with full criteria used for FAST subsample
-- http://casjobs.sdss.org/CasJobs/SubmitJob.aspx
SELECT p.objID, p.ra,p.dec, 
    p.psfMag_r-p.extinction_r + 5*log(sqrt(u.pmRa*u.pmRa + u.pmDec*u.pmDec)/1000.) + 5 AS rpm, 
    p.psfMag_g-p.extinction_g-(p.psfMag_r-p.extinction_r) AS gmr, 
    p.psfMag_r-p.extinction_r-(p.psfMag_i-p.extinction_i) AS rmi, 
    p.psfMag_u, p.psfMag_g, p.psfMag_r, p.psfMag_i, p.psfMag_z, 
    p.psfMagErr_u, p.psfMagErr_g, p.psfMagErr_r, p.psfMagErr_i, p.psfMagErr_z, 
    p.extinction_u, p.extinction_g, p.extinction_r, p.extinction_i, p.extinction_z,
    p.run,p.rerun,p.camcol,p.field,p.obj, p.status,p.nChild,p.flags, 
    u.pmRa, u.pmRaErr, u.pmDec, u.pmDecErr, u.dist20, u.dist22,
    ISNULL(s.specObjID,0) as specObjID, ISNULL(s.specClass,0) as specClass, 
    ISNULL(s.z,0) as z, ISNULL(s.zConf,0) as zConf, ISNULL(s.zWarning,0) as zWarning, 
    ISNULL(s.plate,0) as plate, ISNULL(s.mjd,0) as mjd, 
    ISNULL(s.fiberID,0) as fiberID 
  FROM  
    PhotoTag p JOIN ProperMotions u ON p.objID = u.objID 
    LEFT OUTER JOIN SpecObjAll s ON p.objID = s.bestObjID 
  WHERE 
    p.dec>-5
    AND p.type = dbo.fPhotoType('Star') 
    AND (p.flags & dbo.fPhotoFlags('EDGE')) = 0 
    AND (p.psfMag_r-p.extinction_r) BETWEEN 15 AND 17
    AND u.match = 1 AND u.sigRa<350 AND u.sigDec < 350 AND u.nFit > 2
    AND sqrt(u.pmRa*u.pmRa + u.pmDec*u.pmDec) > 40
    AND ( abs(u.pmRa/u.pmRaErr) > 3 OR abs(u.pmDec/u.pmDecErr) > 3 )
    AND ( (p.psfMag_g-p.extinction_g)-(p.psfMag_r-p.extinction_r) )>0.9
    AND ((p.psfMag_r-p.extinction_r)- (p.psfMag_i-p.extinction_i)) 
         < (0.63*( (p.psfMag_g-p.extinction_g)-(p.psfMag_r-p.extinction_r) ))-0.3362
    AND ((p.psfMag_r-p.extinction_r)- (p.psfMag_i-p.extinction_i)) 
         > (0.3556*( (p.psfMag_g-p.extinction_g)-(p.psfMag_r-p.extinction_r) ))-0.17
}
\end{verbatim}
}

\end{document}

%% file: defs.tex
\newcommand{\aox}{\ifmmode{\alpha_{\mathrm{ox}}} \else $\alpha_{\mathrm{ox}}$\fi} 
\newcommand{\atoms}{\ifmmode{\mathrm{\,atoms~cm^{-2}}} \else \,atoms cm$^{-2}$\fi}
\newcommand{\ax}{\ifmmode{\alpha_x} \else $\alpha_x$\fi} 
\newcommand{\cmsq}{\ifmmode{\mathrm{cm^{-2}}} \else cm$^{-2}$\fi}
\newcommand{\degs}{\ifmmode ^{\circ}\else$^{\circ}$\fi}
\newcommand{\degsq}{\ifmmode {\mathrm{deg^2}} \else deg$^2$\fi}
\newcommand{\perdegsq}{\ifmmode {\mathrm{deg^{-2}}} \else deg$^{-2}$\fi}
\newcommand{\ew}{\ifmmode{W_{\lambda}} \else $W_{\lambda}$\fi}
\newcommand{\fcgs}{\ifmmode erg~cm^{-2}~s^{-1[B}\else erg~cm$^{-2}$~s$^{-1}$\fi}
\newcommand{\flamcgs}{\ifmmode erg\,cm^{-2}\,s^{-1}\,\AA^{-1}\else erg\,cm$^{-2}$\,s$^{-1}$\,\AA$^{-1}$)\fi}
\newcommand{\fnucgs}{\ifmmode {\mathrm{erg~cm^{-2}~s^{-1}~Hz^{-1}}}\else erg~cm$^{-2}$~s$^{-1}$~Hz$^{-1}$\fi}
\newcommand{\gax }{{\lower0.8ex\hbox{$\buildrel >\over\sim$}}}
\newcommand\ha{\ifmmode {\mathrm H}\alpha \else H$\alpha$\fi}
\newcommand\hb{\ifmmode {\mathrm H}\beta \else H$\beta$\fi}
\newcommand{\kms}{\ifmmode~{\mathrm{km~s}}^{-1}\else ~km~s$^{-1}~$\fi}
\newcommand{\lax }{{\lower0.8ex\hbox{$\buildrel <\over\sim$}}}
\newcommand{\lcgs}{\ifmmode erg~s^{-1}\else erg~s$^{-1}$\fi}
\newcommand{\lnucgs}{\ifmmode erg~s^{-1}~Hz^{-1}\else erg~s$^{-1}$~Hz$^{-1}$\fi}

\newcommand{\logz}{\ifmmode{\mathrm{log}}~z \else log$~z$\fi}
\newcommand{\lo}{\ifmmode l_o \else $~l_o$\fi}
\newcommand{\lx}{\ifmmode l_x \else $~l_x$\fi}
\newcommand{\lbol}{\ifmmode L_{\mathrm bol} \else $L_{\mathrm bol}$\fi}
\newcommand{\mone}{\ifmmode ^{-1}\else$^{-1}$\fi}
\newcommand{\msun}{\ifmmode {M_{\odot}}\else${M_{\odot}}$\fi}
\newcommand{\mtwo}{\ifmmode ^{-2}\else$^{-2}$\fi}
\newcommand{\nhgal}{\ifmmode{ N_{H}^{Gal}} \else N$_{H}^{Gal}$\fi}
\newcommand{\nh}{\ifmmode{\mathrm N_{H}} \else N$_{H}$\fi}
\newcommand{\oi}{\ifmmode{\mathrm [O\,II]} \else [O\,II]\fi}
\newcommand{\oii}{\ifmmode{\mathrm [O\,II]} \else [O\,II]\fi}
\newcommand{\oiii}{\ifmmode{\mathrm [O\,III]} \else [O\,III]\fi}
\newcommand{\XMM}{XMM-{\em Newton}}
\def\geqsim{\lower.73ex\hbox{$\sim$}\llap{\raise.4ex\hbox{$>$}}$\,$}
\def\leqsim{\lower.73ex\hbox{$\sim$}\llap{\raise.4ex\hbox{$<$}}$\,$}

\newcommand{\umg}{\ifmmode{\mathrm{(}u-g\mathrm{)}} \else ($u-g$)\fi}
\newcommand{\gmr}{\ifmmode{\mathrm{(}g-r\mathrm{)}} \else ($g-r$)\fi}
\newcommand{\rmi}{\ifmmode{\mathrm{(}r-i\mathrm{)}} \else ($r-i$)\fi}
\newcommand{\gmi}{\ifmmode{\mathrm{(}g-i\mathrm{)}} \else ($g-i$)\fi}
\newcommand{\imz}{\ifmmode{\mathrm{(}i-z\mathrm{)}} \else ($i-z$)\fi}
\newcommand{\jmh}{\ifmmode{\mathrm{(}J-H\mathrm{)}} \else ($J-H$)\fi}
\newcommand{\hmk}{\ifmmode{\mathrm{(}H-K\mathrm{)}} \else ($H-K$)\fi}
\newcommand{\Teff}{\ifmmode{T_{\rm eff}} \else $T_{\rm eff}$\fi}
\newcommand{\ctwo}{\ifmmode C_2 \else C$_2$\fi}

%% file: FHLC0.tex
\begin{deluxetable}{lrccccccccccrrcl}
\tablenum{1}
\tabletypesize{\scriptsize}
\tablecaption{Carbon Stars with SDSS Spectroscopy\label{tab:fhlc}}
\tablewidth{0pt}
\tablehead{
\colhead{Name SDSS J+\,\tablenotemark{a}} & 
\colhead{Epoch\,\tablenotemark{b}} & 
\colhead{{\it r}} & 
\colhead{{\it u-g}} & 
\colhead{{\it g-r}}&
\colhead{{\it r-i}} &
\colhead{{\it i-z}} &
\colhead{J-H\,\tablenotemark{c}} &
\colhead{H-K\,\tablenotemark{c}} & 
\colhead{r-J} &
\colhead{NUV} &
\colhead{FUV} &
\colhead{$\mu_{\alpha}$} &
\colhead{$\mu_{\delta}$} &
\colhead{$M_{\rm i}$\,\tablenotemark{e}} &
\colhead{Class\,\tablenotemark{f}}  \\
  & & & & & & & & & & & & \multicolumn{2}{c}{(mas\,yr$^{-1}$)\,\tablenotemark{d}} & & }
\startdata
000015.01+281009.4 & 52911.21 & 18.10 & 1.63 & 0.81 & 0.25 & 0.18 & 0.72 & 0.38 &  1.26 & \ldots & \ldots & $ 10.5\pm3.0$ & $ -9.8\pm3.0$ & 7.24 & d \\ 
000035.58-001115.0 & 52231.20 & 18.80 & 2.78 & 1.46 & 0.48 & 0.22 & 0.59 & 0.24 &  1.85 & \ldots & \ldots & $  5.7\pm4.3$ & $-65.4\pm4.3$ & 8.40 & d \\ 
000324.63+243815.2 &  \ldots & 17.97 & 1.28 & 0.73 & 0.21 & 0.17 & 0.20 & 0.77 &  1.58 & 23.34 & \ldots & $  6.5\pm2.6$ & $ -1.4\pm2.6$ & 6.80 & uG \\ 
000354.23-104158.2 & 51792.43 & 18.60 & 4.18 & 1.39 & 0.46 & 0.24 & 0.69 & 0.03 &  1.93 & \ldots & \ldots & $ 37.0\pm3.7$ & $-13.9\pm3.7$ & 8.33 & d \\ 
000457.12+010937.9 & 52231.20 & 18.39 & 2.54 & 1.17 & 0.39 & 0.25 & 0.55 & 0.03 &  1.64 & \ldots & \ldots & $ 20.7\pm3.4$ & $-38.9\pm3.4$ & 8.15 & d \\ 
000643.14+155800.7 & 52170.29 & 19.76 & 1.92 & 1.75 & 0.47 & 0.14 & 0.68 & 0.52 &  1.93 & \ldots & \ldots & $-13.0\pm5.1$ & $-38.3\pm5.1$ & 8.32 & d \\ 
000712.54+010757.9 &  \ldots & 18.57 & 1.76 & 0.79 & 0.26 & 0.12 & 0.62 & -0.37 &  1.26 & 24.21 & \ldots & $  3.2\pm3.4$ & $ -2.9\pm3.4$ & 7.40 & u \\ 
001145.30-004710.2 &  \ldots & 18.10 & 2.93 & 1.52 & 0.54 & 0.30 & 0.68 & 0.30 &  1.92 & \ldots & \ldots & $ -4.2\pm3.0$ & $  1.5\pm3.0$ & 8.50 & u \\ 
001245.81-010522.0 & 52910.30 & 19.55 & 2.93 & 1.64 & 0.48 & 0.18 & 0.74 & 0.49 &  1.80 & \ldots & \ldots & $  4.6\pm4.7$ & $-49.9\pm4.7$ & 8.39 & d \\ 
001656.41+010549.8 &  \ldots & 17.85 & 3.05 & 1.27 & 0.45 & 0.44 & 0.58 & 0.26 &  2.17 & \ldots & \ldots & $  0.5\pm3.2$ & $  1.6\pm3.2$ & 8.33 & u \\ 
001716.46+143840.8 & 52170.30 & 18.34 & 2.42 & 1.58 & 0.51 & 0.25 & 0.69 & 0.30 &  1.88 & \ldots & \ldots & $ 16.2\pm3.2$ & $-22.8\pm3.2$ & 8.46 & d \\ 
001836.23-110138.5 &  \ldots & 18.68 & 2.21 & 0.97 & 0.28 & 0.20 & \ldots & \ldots & \ldots & \ldots & \ldots & $  8.5\pm3.8$ & $  1.5\pm3.8$ & 7.49 & u \\ 
002011.85+002912.2 &  \ldots & 15.76 & 2.50 & 1.03 & 0.27 & 0.21 & 0.55 & 0.13 &  1.63 & 23.03 & \ldots & $ -5.5\pm2.5$ & $ -5.9\pm2.5$ & 7.50 & g \\ 
002347.62+000528.9 &  \ldots & 16.81 & 2.52 & 1.05 & 0.23 & 0.24 & 0.51 & 0.09 &  1.64 & 24.36 & \ldots & $  1.0\pm2.7$ & $ -3.3\pm2.7$ & 7.23 & u \\ 
002714.21+002914.3 &  \ldots & 20.74 & 2.11 & 1.86 & 0.54 & 0.28 & 1.04 & 0.22 &  1.98 & \ldots & \ldots & \ldots & \ldots & 8.59 & nL \\ 
002919.44+004314.2 & 52908.30 & 18.89 & 1.34 & 0.74 & 0.23 & 0.12 & \ldots & \ldots & \ldots & 24.25 & \ldots & $ 12.7\pm3.3$ & $  0.2\pm3.3$ & 7.25 & d \\ 
003013.10-003226.6 &  \ldots & 19.50 & 2.01 & 1.88 & 0.60 & 0.16 & 0.77 & 0.63 &  2.10 & \ldots & \ldots & $  7.1\pm4.7$ & $ -7.7\pm4.7$ & 8.90 & u \\ 
003351.80+151114.3 &  \ldots & 20.61 & 1.59 & 1.74 & 0.56 & 0.31 & 1.50 & 0.45 &  1.16 & \ldots & \ldots & $  4.3\pm5.4$ & $  0.8\pm5.4$ & 8.56 & uL \\ 
003504.79+010845.9 &  \ldots & 17.85 & 3.53 & 1.98 & 0.80 & 0.45 & 0.85 & 0.50 &  3.23 & 23.52 & \ldots & $ -0.5\pm2.9$ & $  3.9\pm2.9$ & 9.74 & uE \\ 
003721.34+001224.7 & 52963.17 & 19.07 & 1.95 & 0.90 & 0.24 & 0.18 & 0.49 & 0.11 &  1.51 & \ldots & \ldots & $ 19.1\pm3.3$ & $-19.6\pm3.3$ & 7.35 & d \\ 
\enddata
\tablenotetext{\ }{NOTE: This table is published in its entirety in the
  online electronic edition of this article. A portion is shown here
  for guidance regarding its form and content.}
\tablenotetext{a}{~~Coordinate names are truncated. rather than
  rounded; precise astrometry is available in the SDSS archive. } 
\tablenotetext{b}{~~Epoch in Modified Julian Day provided for those
  objects with significant proper motions as described in the text.}
\tablenotetext{c}{~~Near-IR magnitudes taken from UKIDSS where  available, then 2MASS PSC.}
\tablenotetext{d}{~~Proper motion errors are $\pm 1\sigma$  uncertainties.}
\tablenotetext{e}{~~$M_{\rm i}$ calculated from ($r-i$) color as
  described in \S\,10, {\em assuming} that the object is a main sequence star.}
\tablenotetext{f}{~~Luminosity Classes (lower case) as described in \S\,9.1 are
 d = Dwarf; 
 g = Giant; 
 u = Uncertain; 
 n = No proper motion data.  
Spectral classes/notes (upper case) are
 G: ``G-type carbon star'';
 E: emission lines;
 L: low S/N spectrum;
 N: likely N-type (extremely red, usually strong CN);
 T: candidate extragalactic object;
 X: X-ray source;
 R: in Draco dwarf galaxy (Draco C-1).
A superscript ``1'' means that the object was already published in
  \citep{Downes04}.}
\end{deluxetable}

%% file: DQ0.tex
\begin{deluxetable}{lrccccccrrrl}
\tablenum{3}
\tabletypesize{\scriptsize}
\tablecaption{DQ White Dwarfs with SDSS Spectroscopy\label{tab:dqs}}
\tablewidth{0pt}
\tablehead{
\colhead{Name SDSS J+\,\tablenotemark{a}} & 
\colhead{Epoch\,\tablenotemark{b}} & 
\colhead{{\it r}} & 
\colhead{{\it u-g}} & 
\colhead{{\it g-r}}&
\colhead{{\it r-i}} &
\colhead{{\it i-z}} &
\colhead{NUV} &
\colhead{FUV} &
\colhead{$\mu_{\alpha}$} &
\colhead{$\mu_{\delta}$} &
\colhead{Notes}  \\
 & & & & & & & & & \multicolumn{2}{c}{(mas\,yr$^{-1}$)\,\tablenotemark{d}} & 
}
\startdata
000705.02+282104.2 & 52909.37 & 19.57 & 0.34 & 0.14 & -0.06 & -0.23 & \ldots & \ldots & $ 43.5\pm4.6$ & $-43.6\pm4.6$ &   \\ 
000807.54-103405.5 & 51814.36 & 18.93 & 0.31 & 0.07 & -0.12 & -0.04 & 20.29 & \ldots & $  0.0\pm0.0$ & $  0.0\pm0.0$ &  W  \\ 
002531.50-110800.8 & 51792.44 & 18.00 & 0.22 & -0.00 & -0.10 & -0.17 & 19.18 & \ldots & $110.7\pm3.1$ & $-72.2\pm3.1$ &  W  \\ 
010647.93+151327.6 & 53243.38 & 19.21 & -0.51 & -0.35 & -0.32 & -0.39 & \ldots & \ldots & $-15.1\pm3.3$ & $-28.7\pm3.3$ &   \\ 
010748.20+010240.1 & 52963.19 & 18.58 & 0.55 & 0.24 & -0.00 & -0.01 & 20.66 & \ldots & $ 20.6\pm2.9$ & $-169.7\pm2.9$ &   \\ 
015441.74+140307.9 & 51465.37 & 17.64 & 0.56 & 0.23 & 0.06 & -0.08 & 20.26 & \ldots & $165.1\pm2.8$ & $-220.8\pm2.8$ &  M \\ 
015655.97-005036.3 & 52224.29 & 18.89 & 0.47 & 0.25 & 0.03 & -0.09 & \ldots & \ldots & $  0.0\pm0.0$ & $  0.0\pm0.0$ &   \\ 
020534.13+215559.7 & 53269.28 & 19.50 & 1.17 & 0.41 & 0.06 &  0.02 & \ldots & \ldots & $137.8\pm3.4$ & $ 36.8\pm3.4$ & C  \\ 
020906.07+142520.7 & 53710.27 & 19.16 & 0.42 & 0.13 & -0.02 & -0.03 & 21.16 & \ldots & $-23.2\pm2.9$ & $-48.4\pm2.9$ &  W  \\ 
022909.51+251024.4 & 55119.44 & 18.84 & 0.31 & 0.04 & -0.05 & -0.12 & 20.46 & \ldots & $ 70.1\pm2.8$ & $-11.8\pm2.8$ &   \\ 
023609.38-080823.9 & 54057.23 & 19.81 & 0.11 & -0.06 & -0.10 & -0.08 & 19.86 & \ldots & $141.6\pm5.0$ & $ 42.2\pm5.0$ &   \\ 
023945.01+002745.0 & 52963.25 & 19.60 & 0.45 & 0.13 & -0.03 & -0.05 & \ldots & \ldots & $129.8\pm3.6$ & $-12.2\pm3.6$ &  W  \\ 
024332.74+010112.3 & 54715.46 & 20.28 & -0.09 & 0.15 & -0.05 & -0.33 & \ldots & \ldots & $ 59.2\pm3.4$ & $-121.5\pm3.4$ &   \\ 
024802.27+340802.4 & 53678.43 & 18.46 & 0.72 & 0.37 & 0.14 & -0.06 & 21.77 & \ldots & $-33.7\pm3.0$ & $-158.7\pm3.0$ &   \\ 
030538.53+055734.3 & 53655.44 & 20.67 & 0.45 & 0.01 & 0.02 & -0.11 & 22.63 & \ldots & $ -1.5\pm5.0$ & $-40.0\pm5.0$ &   \\ 
032054.11-071625.4 & 51465.45 & 19.32 & 0.23 & 0.43 & 0.12 &  0.06 & 21.09 & \ldots & $123.6\pm3.9$ & $-11.7\pm3.9$ &  P E \\ 
033218.22-003722.1 & 53270.42 & 18.37 & 0.21 & 0.02 & -0.06 & -0.11 & 19.62 & \ldots & $ -0.9\pm5.4$ & $-173.5\pm5.4$ &   \\ 
041601.25+071308.9 & 54040.43 & 19.34 & 0.39 & 0.21 & -0.01 & -0.16 & 21.15 & \ldots & $ 11.8\pm3.3$ & $ 22.4\pm3.3$ &   \\ 
045032.77-002959.5 & 53270.47 & 19.78 & -0.01 & 0.14 & -0.01 & -0.10 & \ldots & \ldots & $  0.0\pm0.0$ & $  0.0\pm0.0$ &  \\ 
073703.83+645524.6 & 53293.47 & 19.52 & 0.37 & 0.11 & 0.00 & -0.15 & 21.20 & \ldots & $ 14.1\pm3.6$ & $-57.7\pm3.6$ &   \\ 
\enddata
\tablenotetext{\ }{NOTE: This table is published in its entirety in the
  online electronic edition of this article. A portion is shown here
  for guidance regarding its form and content.}
\tablenotetext{a}{~~Coordinate names are truncated.
rather than rounded; precise astrometry is available in the SDSS archive.}
\tablenotetext{b}{~~Epoch in Modified Julian Day provided for those
  objects with significant proper motions as described in the text.}
\tablenotetext{d}{~~Proper motion errors are $\pm 1\sigma$ uncertainties.}
\tablenotetext{f}{~~Notes on spectrum are:
 C = unusual continuum shape and/or possible composite;
 E = emission lines;
 L = low S/N spectrum;
 M = detected in 2MASS;
 P = deep molecular bandheads;
 W = weak molecular bandheads
}
\end{deluxetable}

%% file: FASTdc0.tex
\begin{deluxetable}{lrcccccccccr}
\tablenum{4}
\footnotesize
\tablecaption{Dwarf Carbon Stars Confirmed with FAST Spectroscopy\label{tab:FASTdcs}}
\tablewidth{0pt}
\tablehead{
\colhead{Name SDSS J+\,\tablenotemark{a}} & 
\colhead{Epoch\,\tablenotemark{b}} & 
\colhead{{\it r}} & 
\colhead{{\it u-g}} & 
\colhead{{\it g-r}}&
\colhead{{\it r-i}} &
\colhead{{\it i-z}} &
\colhead{J-H} &
\colhead{H-K} & 
\colhead{r-J} &
\colhead{$\mu_{\alpha}$} &
\colhead{$\mu_{\delta}$}\\
 & & & & & & & & & & \multicolumn{2}{c}{(mas yr$^{-1}$)\,\tablenotemark{c}}
} 
\startdata
015856.70+141236.8 & 51464.37 & 15.96 & 2.53 & 1.25 & 0.38  & 0.20 & 0.53 & 0.05 &  1.54 & $110.6\pm2.7$ & $-27.5\pm2.7$ \\
081913.04+023724.0 & 52318.23 & 16.16 & 3.31 & 1.72 & 0.00  & 0.28 & 0.80 & 0.38 &  2.11 & $-41.6\pm2.5$ & $-46.2\pm2.5$ \\
083451.15+074008.7 & 52667.34 & 16.56 & 2.85 & 1.71 & 0.00  & 0.36 & 0.67 & 0.32 &  2.14 & $ 49.5\pm2.8$ & $-39.8\pm2.8$ \\
093334.14+064812.6 & 52338.17 & 15.32 & 2.67 & 1.28 & 0.43  & 0.25 & 0.59 & 0.11 &  1.74 & $ -9.1\pm2.5$ & $-43.4\pm2.5$ \\
093547.66+270938.9 & 53111.21 & 15.62 & 2.49 & 1.10 & 0.33  & 0.16 & 0.51 & 0.15 &  1.58 & $ 17.6\pm2.5$ & $-93.5\pm2.5$ \\
145725.86+234125.5 & 53148.34 & 16.77 & 2.81 & 1.79 & 0.00  & 0.29 & \ldots & \ldots & \ldots & $-360.6\pm2.6$ & $-68.1\pm2.6$ \\
151040.69+055949.4 & 52076.21 & 16.35 & 2.65 & 1.27 & 0.00  & 0.24 & 0.60 & 0.14 &  1.84 & $-67.2\pm12.3$ & $-110.6\pm12.3$ \\
151702.65+152926.2 & 53439.48 & 16.06 & 2.37 & 1.03 & 0.00  & 0.15 & 0.57 & 0.10 &  1.53 & $-58.8\pm2.4$ & $-11.8\pm2.4$ \\
164035.59+125208.8 & 53148.41 & 15.40 & 2.80 & 1.36 & 0.38  & 0.18 & \ldots & \ldots & \ldots & $-141.4\pm2.6$ & $-324.3\pm2.6$ \\
184735.67+405944.2 & 53532.40 & 15.68 & 2.84 & 1.95 & 0.83  & 0.51 & 0.70 & 0.36 &  2.70 & $ 54.5\pm2.7$ & $-46.7\pm2.7$ \\
\enddata
\tablenotetext{a}{~~Coordinate names are truncated
rather than rounded; precise astrometry is available in the SDSS archive.}
\tablenotetext{b}{~~Epoch in Modified Julian Day provided for those
  objects with significant proper motions as described in the text.}
\tablenotetext{c}{~~Errors are $\pm 1\sigma$ uncertainties}
\end{deluxetable}